%% file: HIM_JJ_v8.tex
\documentclass[aps,prl,twocolumn,groupedaddress]{revtex4-2}

\usepackage{lipsum}
\usepackage{graphicx, caption, subcaption}
\usepackage{siunitx}
\usepackage{tikz}
\usepackage{float}
\usepackage{physics}
\usepackage{mathtools}
\usepackage{amssymb}

\usepackage{booktabs} 

\begin{document}


\title{Highly tunable NbTiN Josephson junctions fabricated with focused helium ion beam}


\author{Aki Ruhtinas}
\email[email: ]{akperuht@jyu.fi}
\author{Ilari J. Maasilta}
\email{maasilta@jyu.fi}

	\affiliation{Nanoscience Center, Department of Physics, University of Jyv{\"a}skyl{\"a}, FI-40014 Jyv{\"a}skyl{\"a}, Finland}%

\date{\today}

\begin{abstract}
 We demonstrate a direct writing method for the fabrication of planar Josephson junctions from high quality superconducting niobium titanium nitride (NbTiN) thin films, by creating local disorder using focused He-ion beam irradiation in a helium ion microscope. 
We show that we can control the suppression of superconductivity in NbTiN as a function of the helium ion beam fluence, 
enabling us to successfully fabricate Josephson junctions with highly tunable weak links 
ranging from metallic to insulating phase, due to the continuous nature of the disorder-induced superconductor-insulator transition. We demonstrate the successful fabrication of both SNS and SIS type of devices, and show that we can achieve exceptionally wide range ($\sim$5 orders of magnitude) of critical current densities and junction resistances. 
The SNS type junctions follow closely the ideal resistively and capacitively shunted junction behavior, have high characteristic voltages up to $\sim$ 1.5\,mV and show Shapiro steps up to very high orders, 
while the SIS junctions also follow  established theories well.  
The results suggest that junctions fabricated with 
this method from NbTiN are suitable for a wide range of applications in superconducting electronics 
because of the excellent mechanical, electrical and microwave properties of NbTiN. In particular, as NbTiN has previously been used to fabricate high quality factor microwave resonators, we see 
the method as 
a promising and simple way to realize superconducting qubits and other quantum devices using only a single superconducting film.

\end{abstract}


\maketitle

\section{Introduction}
While weak disorder does not have an effect on the superconducting state \cite{Anderson1959}, strong enough disorder can cause a superconductor to undergo a superconductor-to-insulator transition (SIT) \cite{Gantmakher2010}. 
If a superconducting film is locally pushed into the strong disorder limit, by ion irradiation for example, the arising SIT quantum phase transition can be exploited in the fabrication of weak links for Josephson junctions from a single material, as changing the level of disorder can change the film properties drastically. However, to make a good junction, the disordered region needs to be defined with nanometer resolution. Ion irradiation or ion implantation of weak links is not new idea\cite{Likharev_weak_links}, but a high enough spatial resolution (beam diameter of $\sim 0.5$\,nm) became only available in 2007 with the commercial introduction of helium ion microscope (HIM). 
In the helium-ion direct-write fabrication of Josephson junctions, this high resolution helium ion beam is used to locally irradiate a superconducting film to introduce disorder and therefore control the SIT, enabling fabrication of highly tunable weak links 
where both the width and the strength of the link can be tuned continuously\cite{Cybart2015}. He-ions are needed, instead of heavier ions such as Ga, to significantly reduce the probability of sputtering, which would lead to milling instead of introduction of disorder. 

He-ion direct writing of Josephson junctions was first demonstrated in YBCO films \cite{Cybart2015}, After that seminal study, the method has been applied not only to YBCO thin films to create Josephson junctions and SQUIDS \cite{Muller2019, LiHao2020, Chen2022}, but also to MgB$_2$ \cite{LKasaei2018} and to Bi$_2$Sr$_2$CaCu$_2$O$_{8+x}$ \cite{YanTing2021} thin films. However, these materials are often 
difficult to work with, and electrical, mechanical and chemical properties are not good enough for some applications. For many superconducting devices, instead of using complex high-temperature superconducting materials, the more standard low-temperature superconductors with more ideal BCS-type behavior are desirable. 
Superconducting nitrides are such materials, and also seem a promising choice for direct writing, as it is known that a disorder driven superconductor-insulator transition exists in NbN \cite{Makise2015,Ezaki2012,Yong2013}, TiN \cite{Hadacek2004,Baturina2007} and NbTiN \cite{Burdastyh2017,Burdastyh2020}.
Recently, direct He-ion writing of NbN has indeed been investigated \cite{Martinez2020,Li2023}.   


Here, we focus on using high quality ($T_c > 15$\,K) pulsed laser deposited (PLD) NbTiN as the superconducting material, to which we direct write Josephson junctions with a nanoscale helium ion beam in a helium ion microscope.  NbTiN has excellent electrical and mechanical characteristics and high corrosion resistance, which makes it a good choice for Josephson junction fabrication. 
Moreover, NbTiN is already a state-of-the-art material in many devices, and it has been used for example in high quality factor microwave resonators {\cite{Barends2010,Bruno2015,Vissers2015,Sullivan2022,Muller_2022feb}, rapid single flux quantum (RSFQ) devices \cite{Yu_2006}, superconducting nanowire single-photon detectors (SNSPDs)\cite{Schuck2013,Esmaeil2017} and in THz-band heterodyne sensing technologies \cite{Karpov2009,Shiba2012}. 

The measured electrical characteristics of the junctions closely follow established theories. Moreover, we show that we can tune both the junction length and the critical current density of the junctions over a large range (five orders of magnitude for critical current density), making these junctions exceptionally tunable and applicable to optimizing the performance of a wide range of devices.  
Particularly appealing feature of this junction fabrication method is the opportunity to fabricate complete quantum devices (for example qubits and parametric amplifiers) using only a single NbTiN thin film, 
as this would reduce the number of interfaces, materials and processing steps to be optimized, known to be critical factors for performance degradation in superconducting devices through various loss mechanisms\cite{Murray2021}.     


\begin{figure*}[t]
		\includegraphics[width=0.7\linewidth]{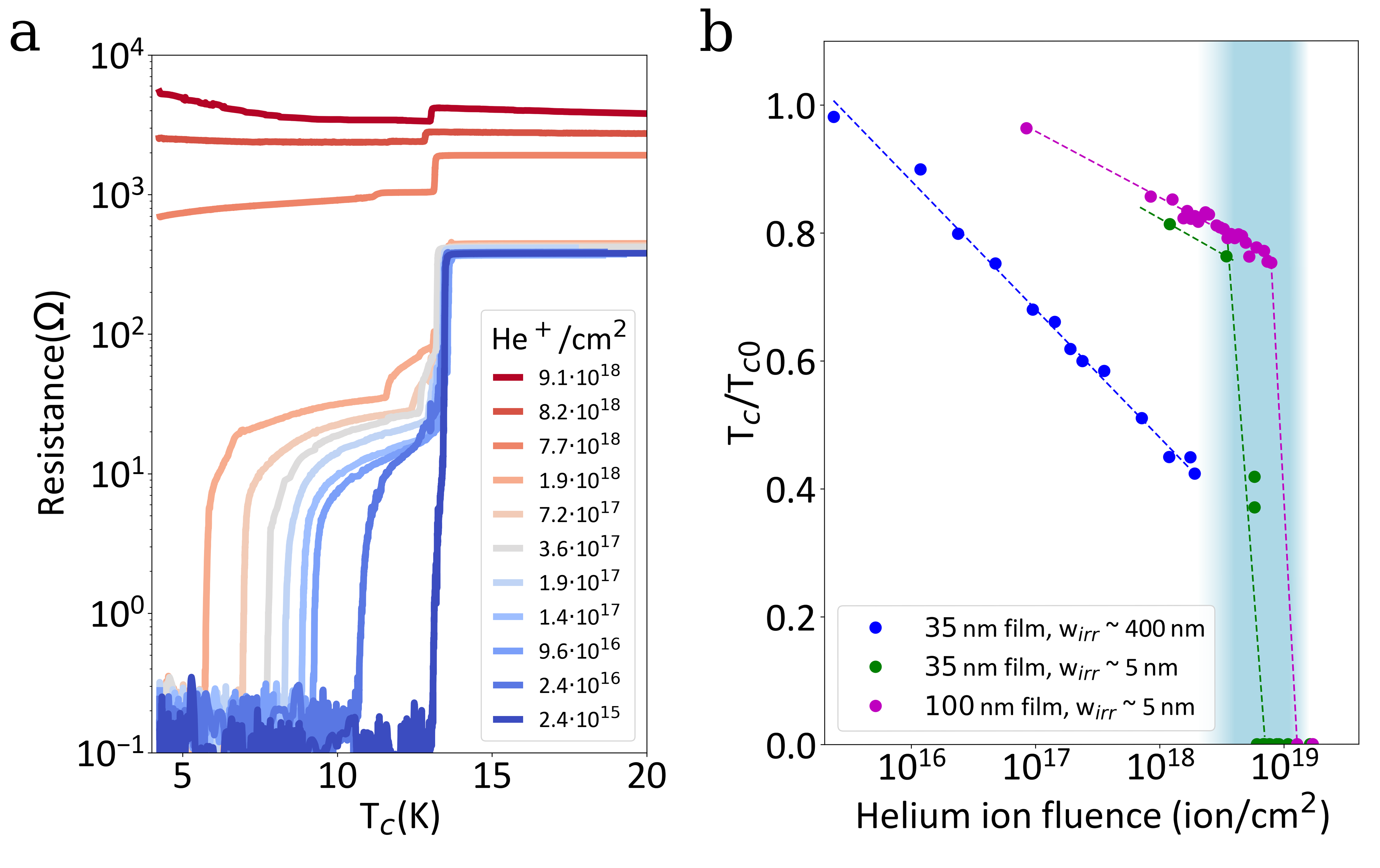}
		\caption{Suppression of $T_c$ and the demonstration of disorder driven SIT in NbTiN thin films. Panel (a) shows the resistances of superconducting NbTiN wires after the irradiation of a region of a wire with a helium ion beam using different He ion beam fluences. Wires undergo a transition from superconducting to insulating phase as the helium ion fluence increases, with a suppression of $T_c$ along the way. Width of irradiated region $w_{irr} \sim 400$ nm ,except for the three highest fluence samples, for which $w_{irr} \sim 5$ nm. All samples are from the same chip, with a film thickness 35 nm. Panel (b) shows the $T_c$s of the irradiated regions of a collection of wires from three chips in a semilogarithmic plot, showing a linear dependence of $T_c$ as a function of the logarithm of the ion beam fluence. SIT region is shaded in blue.}
		\label{fig:tc_suppression}
\end{figure*}

\section{Device fabrication}
\subsection{Film deposition and lithography}
Thin films of 
NbTiN were deposited using the same reactive infrared pulsed laser deposition (PLD) technique as demonstrated for TiN \cite{Torgovkin_2018}, TaN\cite{Chaudhuri2013} and NbN\cite{Chaudhuri2011} before. As a target, we used two different compositions, Nb$_{0.39}$Ti$_{0.61}$ and Nb$_{0.5}$Ti$_{0.5}$, with most of the samples made from the 50/50 target. The composition choice did not have a significant effect on the film properties, with only a small effect on the resulting NbTiN film resistivity.  
 We deposited the films in ultra high purity nitrogen atmosphere using $700\,^{\circ}$C substrate temperature and cubic (100)-oriented MgO substrates. MgO substrates were chosen for several reasons. First of all, MgO has a good lattice match to NbTiN, allowing epitaxial growth. Second, the SF$_6$-based plasma etching step to define the wires stops to MgO. Third and most importantly, MgO seems to handle large helium ion fluences without issues with the formation of helium bubbles in the substrate, unlike what could sometimes happen in Si, for example \cite{Zhang_2015}.
However, we want to stress here that MgO is not the only possible substrate material.   

The superconducting properties of our PLD deposited films are excellent, and with 100\,nm thick films the superconducting critical temperature $T_c$ is as high as 15.7\,K and resistivity is as low as $\sim$ \SI{30}{\micro\ohm\centi\meter},
with the superconducting transition width typically quite narrow ($\sim$50\,mK). The critical temperature decreases when the film thickness decreases, and for a 50\,nm thickness $T_c \sim$15\,K  while for a 22\,nm thick film, we have achieved a $T_c$ of 11.5\,K. In this study, we have mostly used a film thickness of 35\,nm, with a $T_c \sim$13\,K.

The bonding pads and superconducting wires were defined to the film using electron beam lithography (EBL) and subsequent reactive ion etching in SF$_6$+O$_2$ plasma. More details of the device fabrication process are given in the Supplemental Information.

\subsection{Helium ion irradiation}
Josephson junctions were defined to the superconducting wire using a helium ion beam "direct writing" technique. For this, we used a $30\,$kV He$^+$ beam in a Zeiss Orion Nanofab helium ion microscope with a nominal spatial resolution of 0.5\,nm. The junction was defined as a narrow line irradiation, perpendicular to the length of the superconducting wire, creating a weak link between the two electrodes via disorder (Figure \ref{fig:junction_images_Jc}). Depending on the spot size and focusing, the width of the irradiated line varied between $\sim$3\,nm-500\,nm. Larger irradiation widths were achieved by using large spot size and by defocusing the beam significantly. Compared to the He-ion beam written YBCO junctions \cite{Cybart2015}, the NbTiN junctions in this study need roughly two orders of magnitude more fluence to suppress superconductivity. Thus, in this work a fluence in the range $10^{16}-10^{20}$ He$^+$/cm$^2$ was used. After the irradiation, every junction was imaged with a quick HIM scan to ensure that the junction was successfully defined, and to check the resulting linewidth (corresponding to the junction length in the parlance of Josephson junctions). 

To quantify the possible partially milled trench depth, we have performed milling experiments with the helium ion microscope and the subsequent atomic force microscope (AFM) measurement. 
Using that data, we have determined an estimate for the milling depth $d_{mill}$ as a function of the fluence $\Phi$, giving an empirical expression 
\begin{equation}
d_{mill}[\text{nm}] \approx 4.06\cdot\left(\Phi[10^{18}\text{ion/cm}^2]\right)^{0.68}.
\label{eq:mill}
\end{equation}
We use Equation \eqref{eq:mill} for thickness corrections to determine more realistic values for the critical current densities $J_c$ and resistivities $\rho$ of the junctions. Derivation of Eq. \eqref{eq:mill} is presented in the Supplementary Information.

In addition to irradiation and milling experiments, we have also simulated the ion irradiation with a simulation code for ion-matter interaction (SRIM)\cite{ZIEGLER2010}. Using SRIM, we simulated a $30\,$kV He$^+$ beam impinging on a 35\,nm NbTiN film on top of a MgO substrate. These simulations show (Supplementary Information) that most helium ions travel through them film, stopping only in the substrate (MgO) after traveling roughly 150\,nm, and that there is only a small lateral straggle when the ions travel through the NbTiN film. The created dislocations are concentrated in very narrow region of a few nm, which is good for an accurate junction definition. 

\begin{figure*}[t]
		\includegraphics[width=0.8\linewidth]{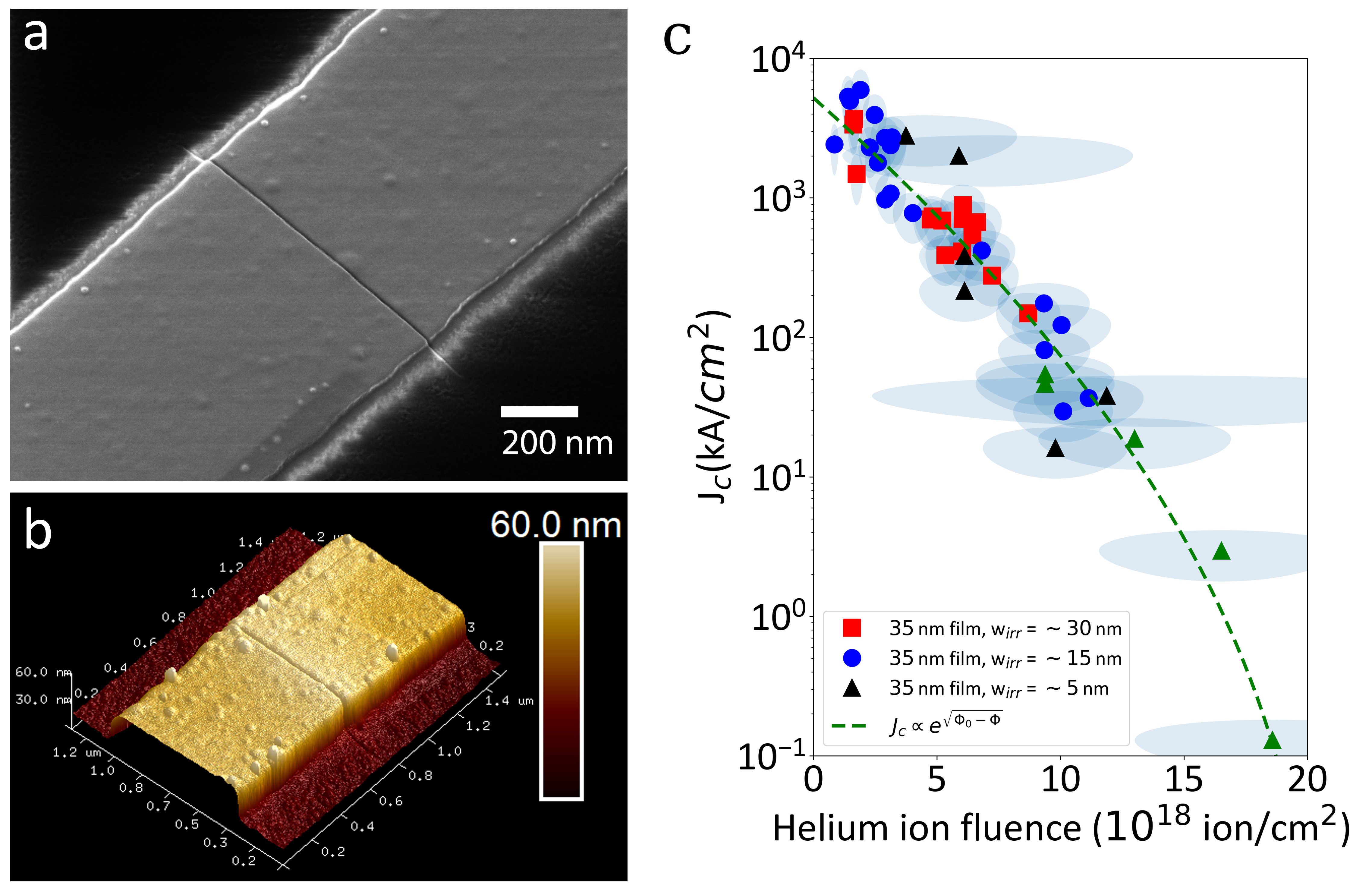}
		\caption{(a) a helium ion micrograph  and (b) an atomic force microscope (AFM) image of a NbTiN Josephson junction created by helium ion irradiation. The fluence used here was $\sim 6\cdot10^{18}$\,ion/cm$^2$. The width of the irradiated region in this case is quite wide, $\sim$14\,nm, and the superconducting wire is partially milled, as can be seen from both the AFM and HIM images. Panel (c) shows a dependence of the critical current density $J_c$ for a set of samples measured at 4.2\,K as a function of the helium ion fluence. Uncertainties are shown as blue shaded regions.}
		\label{fig:junction_images_Jc}
\end{figure*}

\section{Results}
\subsection{Suppression of superconductivity}
In order to investigate the effects of helium ion irradiation on NbTiN thin films first more generally, we have irradiated superconducting NbTiN wires with HIM, using a varying fluence and size (linewidth) of the irradiated region. For small linewidths ($\sim 5$\,nm) we used  spot sizes 4-6, whereas for large linewidths a spot size of 2 was used. Results are shown in Figure \ref{fig:tc_suppression} (a), revealing that helium ion beam irradiation induces strong enough local disorder to suppress superconductivity and ultimately push the irradiated region through the superconductor-insulator transition with increasing fluence. The superconducting critical temperature suppression as a function of the helium ion fluence $\Phi$ is surprisingly linear in semilogarithmic scale (Fig. \ref{fig:tc_suppression} (b)), indicating that $T_c \propto -\log (\Phi)$. This relation allows straightforward and accurate $T_c$ control of the NbTiN film, which may be useful in different applications. We should note that while it is difficult to determine the exact width of a large area irradiation (and thus the precise value of the fluence), the width fluctuations between different irradiation rounds are small, and a possible systematic error would only shift this linear dependency. 

The insulating state arising from this superconductor-insulator transition is interesting in its own right. The resistance of the insulating barrier can increase exponentially over several orders of magnitude from room temperature to cryogenic temperatures. This strong temperature dependence can be explained by variable range hopping theories, indicating that in the insulating state electrons are localized due to strong disorder. When junction resistance as a function of temperature is fitted against variable range hopping theory (Supplementary Information), we find that the Efros and Shklovskii variable range hopping theory \cite{Efros1975} fits well to experimental data. Thus we conclude that as expected, in the insulating side of the transition electrons in the NbTiN film are localized.

While the irradiations with small widths ($\sim 5$\,nm) also have the property $T_c \propto -\log (\Phi)$ at lower fluences, the measured $T_c$ does not reflect the inherent $T_c$ of the irradiated region due to the presence of Josephson supercurrent, but rather the temperature of the disappearance of the supercurrent.  
Nevertheless, from the apparent $T_c$ we can see that with the small linewidth irradiations, there is a sharp transition to the insulating phase beginning around $\sim 2\cdot10^{18}$\,ion/cm$^2$ for 35\,nm thick films, and this transition shifts to a higher fluence ($\sim 1\cdot10^{19}$\,ion/cm$^2$) for 97\,nm films (Fig. \ref{fig:tc_suppression} (b)). This shift is expected, as there are more atoms to displace when films are thicker. However, there is a lot of energy left in He ions after traveling through the NbTiN, so that a film thickness increase does not increase the transition fluence linearly. We find that this SIT region (blue region in Figure \ref{fig:tc_suppression} (b)) is the best for Josephson junction fabrication, as in that case the critical current can be tuned continuously from $ \sim I_c$ of the bulk wire down to zero.

\subsection{Josephson junctions}
After the characterization of the SIT as a function of fluence, we proceeded to fabricate several weak links with varying strengths and widths using fluences in the aforementioned transition region. Figures \ref{fig:junction_images_Jc} (a) and (b)  show an example of microscope images of a fabricated junction, where the junction is formed by the perpendicular irradiation line in the superconducting wire. To be able to reproducibly fabricate junctions, we have determined experimentally how the critical current density $J_c$ depends on the helium ion fluence within the SIT region in Fig. \ref{fig:tc_suppression} (b). The results of these experiments are shown in Figure \ref{fig:junction_images_Jc}(c), where we plot the critical current densities $J_c$ of several junctions fabricated to $\sim 500$\,nm wide and $35$\,nm thick wires. We note that in the determination of $J_c$, we have taken into account the effect of the change of the effective junction area by partial milling, which was determined with AFM. 
 As the fluence was calculated using the FWHM of the junction widths in HIM images, in some cases the error in fluence determination is significant. 
We find that $J_c$ is roughly exponentially proportional to the helium ion fluence, so that $-\log(J_c) \propto \Phi$. 

To understand this behavior in more detail, we note that the relevant length scales for 
 the weak link are the mean free path $\ell$ and the normal metal coherence length $\xi_N$ in the irradiated normal metal region, with $\xi_N$ is given by \cite{Likharev_weak_links}
\begin{equation}
\xi_N(T) = \sqrt{\frac{\hbar D}{2\pi k_BT}},
\label{eq:ksii}
\end{equation}
where $D$ is the diffusion coefficient. As a first approximation, we can assume that helium ion irradiation induces disorder linearly such that $1/k_F\ell \propto \Phi$, where $k_F\ell$ is the disorder parameter with $k_F$ the Fermi wave vector. As the diffusion coefficient is also a linear function of $k_f\ell$, from Eq. \eqref{eq:ksii} we can see that with this assumption $\xi_N \propto 1/\sqrt{\Phi}$. The critical current density in SNS Josephson junctions follows approximately exponential dependence $J_c \propto e^{-L/\xi_N}$ \cite{Likharev_weak_links}, where $L$ is the junction length. Now if we fix the junction length, we can thus explain the observed approximately exponential decay of $J_c$  
with fluence. 
 We note that while this equation is expected to be a good estimation for lower fluences, the simple exponential dependence breaks down at higher fluences, where the weak link is in the insulating phase. In Figure \ref{fig:junction_images_Jc}(c) we have fitted the experimental data with a function $J_c \propto e^{-\sqrt{\Phi_0-\Phi}}$ , where $\Phi_0$ corresponds to the maximum fluence after which the weak link is fully insulating. The fit to the experimental data is reasonably good with $\Phi_0 = 20\cdot10^{18}$\,ion/cm$^2$, and more importantly, this phenomenological equation $J_c \propto e^{-\sqrt{\Phi_0-\Phi}}$ correctly reproduces the behaviour from small to large fluences. 
The good fit to the experimental data gives evidence to the original assumption of 
the disorder increasing approximately linearly with fluence. 

Because the SIT region is relatively narrow in $\Phi$, precise control of the helium ion beam is needed for reproducible fabrication of junctions. This is especially true for the narrowest $<10$\,nm junctions (black triangles in Fig. \ref{fig:junction_images_Jc}(c)), as small changes in the irradiation width change $\Phi$ and thus $I_c$ drastically. From large area irradiations we can see that there is certainly strong suppression of $T_c$ without substantial ion beam milling of the sample, but with the narrow irradiations the required ion fluences are so high that there is substantial milling, as seen from the AFM data (Supplementary Information). 
 However, we estimate that in all of the SNS-type junctions there is still over 10\,nm left. In pristine 10\,nm NbTiN film $T_c$ is as high as $\sim$7\,K, indicating that the suppression of $T_c$ and $I_c$ is not caused by milling 
but by ion beam induced disorder. 
 
 While the fabricated SIS junctions are milled more, we estimate that there is at least $\sim 5$\,nm left, an amount that would still be superconducting if in pristine state and thus also in this case the suppression is caused by ion beam induced disorder and not by simple thinning of the film. 

\begin{figure*}[t!]

\includegraphics[width=\linewidth]{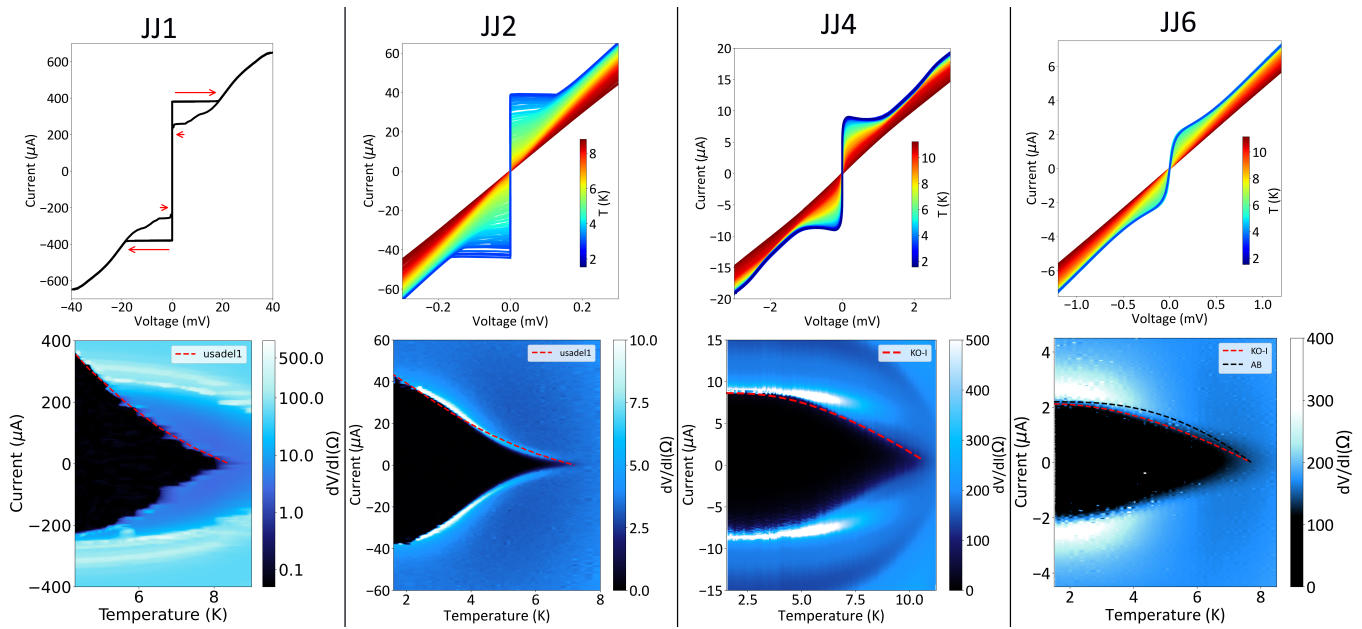}

\caption{DC transport measurements of four representative Josephson junction devices: JJ1, JJ2,JJ4 and JJ6, fabricated using different helium ion fluences and different superconducting wire geometries, listed in Table I. Top row: current-voltage characteristics, bottom row: differential conductance as a function of temperature and current.  For all junctions, comparison to different theories are shown in the right column by the dashed lines. Usadel1 = numerical solution of Usadel equations, KO-I = Kulik and Omelyanchuk theory and AB = Ambegaokar-Baratoff theory, as explained in the text.}

\label{fig_ivc}
\end{figure*}

\subsubsection{DC transport measurements}
Josephson junctions fabricated with helium ion direct writing were characterized via DC transport measurements either down to 1.5\,K with a liquid helium evaporation cryostat or down to 4.2\,K with liquid helium bath. We measured the current-voltage characteristics (IVC) at different temperatures using a battery powered sweep box and a voltage divider, to achieve a low noise level. Voltage and current were measured using Ithaco preamplifiers (models 1201 and 1211), and the differential conductance was determined with numerical differentiation of the IVC. In Figures \ref{fig_ivc} and \ref{fig:JJ7_ivc} we show DC transport measurements of five representative Josephson junctions fabricated with direct writing method. In Figure \ref{fig_ivc} there are transport measurements of four SNS-type junctions, and in Figure \ref{fig:JJ7_ivc} there are transport measurements of an SIS-type of junction. As the junction is continuously tunable from superconducting to insulating state, this division can be somewhat artificial, but in most of the cases clear distinction can be made based on the current-voltage characteristics.

In order to extract more precisely the critical current ($I_c$) and normal state resistance ($R_n$) values, the IVCs of SNS-type of junctions were fitted at selected temperatures using the RCSJ model \cite{tinkham_book} (see the Supplementary Information for more details). From these fits, we extracted values for $I_c$, $R_n$ and  the junction capacitance $C$. However, because self-heating effects quite closely resemble capacitive effects \cite{Likharev_weak_links,Courtois2008}, we used capacitance merely as fit parameter as it does not correspond to the estimated geometrical junction capacitance. Thus we believe that the hysteresis seen in junction JJ1 is most likely of thermal origin instead of capacitive. In addition, because JJ7 and JJ8 have low $I_c$ and were measured only down to 4\,K, thermal noise can suppress the supercurrent significantly, and thus we estimated it from $I_c \approx I_g\pi/4$, where $I_g$ is the quasiparticle 
current inflection point \cite{Sun2016}. The experimental parameters of these junctions are presented in Table I.

The junction length and the milling depth reported in Table I are more challenging to  estimate accurately experimentally, however 
they can be estimated from HIM and AFM images with reasonable accuracy, as described in the Supplementary Information.   
In the case of junction JJ5, the estimation of fluence has exceptionally high uncertainty as there was a shift in the beam. In these cases, we estimate that 50\% fluence is lost to the beam shift, and we have noticed that this is a good estimation in most of the similar cases (based on $J_c$ vs fluence dependence shown Figure \ref{fig:tc_suppression}(c)). 

From the transport measurements we concluded that it is possible to tune the critical current from $I_c$ of the superconducting leads continuously down to zero. Here we show that $I_c$ can be controlled by five orders of magnitude, as in junctions JJ1-JJ8, $I_c$ varies from \SI{390}{\micro\ampere}(JJ1) to \SI{9}{\nano\ampere}(JJ8). The junction dimensions such as the film thickness $t$, the superconducting wire width $W$ and the junction length $L_j$ (=irradiation width) vary between junctions JJ1-JJ8, showing flexibility of the method. JJ4 was fabricated from a 100\,nm thick film, demonstrating that direct writing method is feasible for such films, thicker than the most commonly used $\sim$30\,nm thickness \cite{Cybart2015,LKasaei2018}. 
In addition, we have demonstrated that fabrication of longer junctions is also possible with the direct write method, as JJ2 is 30\,nm long. For that case, the resistivity of the barrier is low, but the long length reduces the critical current. 

\subsubsection{Weak link properties}

When milling was taken into account, the calculated junction resistivities for SNS-type devices ranged from $\sim$\SI{900}{\micro\ohm\centi\meter} (JJ2) to $\sim 1.5\cdot 10^{5}$\SI{}{\micro\ohm\centi\meter} (JJ5). 
These are definitely high values for metals, and therefore the weak links with strongest disorder are in the limit of not being good metals anymore. 
The sheet resistances 
vary between $\sim$0.3\,k$\Omega$ (JJ1, JJ2) to $\sim$110\,k$\Omega$ (JJ5). In all of the SNS type junctions, the temperature dependence of $R_n$ is very weak, indicating still metallic-like conductivity. In NbTiN, a Bose metallic state has been experimentally observed when the sheet resistance $R_{\square}$ is above the superconducting resistance quantum $R_q = h/4e^2 = 6.45$\,k$\Omega$ \cite{Burdastyh2020}. From the measured $R_{\square}$ values, it seems that junctions JJ3-JJ6 are in the vicinity or above this limit, suggesting a possible presence of the Bose metal phase. JJ5 and JJ6 have high enough $R_{\square}$ that based on  Ref. \cite{Burdastyh2017}, these junctions should already be in the insulating side of the SIT, indicating that they could possibly be overdamped SIS junctions. This division is however somewhat artificial as weak links are continuously tunable.

For the SIS junctions JJ7 and JJ8, the calculated barrier resistivities (with milling estimation taken into account) were $\sim 0.02$\,\SI{}{\ohm\meter} and $\sim 3$\,\SI{}{\ohm\meter}, respectively. 
Interestingly, these resistivities have values corresponding to the barrier being in between a good metal and a good insulator.  
As the conductance of the barrier is still relatively high for a tunnel junction, these devices have lower barrier heights than usual SIS devices. To quantify this, 
we extracted the barrier heights and lengths by fitting the Simmons model \cite{Simmons1963} to the higher voltage (up to $\sim$ 50\,mV) conductance data at 20\,K temperature.  The barrier heights for junctions JJ7 and JJ8 are close to each other (both $\sim35$ meV), almost two orders of magnitude lower than the for SIS junctions with aluminium oxide barriers. 
From these fits we can also extract the barrier length (thickness), and for junctions JJ7 and JJ8 these are 2.6\,nm and 3.8\,nm, corresponding well to the values determined from HIM images (Table I). 
We have extracted similar results from other SIS junctions fabricated with the same way, so we can confidently state that this fabrication method creates SIS junctions with low ($\phi_0<$100\,meV) and wide ($L_j\sim$3\,nm) barriers.

To gain more insight to the SIS junction properties, we also determined the normal state tunneling resistance $R_n$, the superconducting gap $\Delta$ and the Dynes broadening parameter $\Gamma$ 
from fits to the quasiparticle tunneling model \cite{tinkham_book}. The determined superconducting gaps are 1.5\,meV (JJ7) and 1.1\,meV (JJ8), substantially lower than the gap in the leads ($\sim 2.4$ meV). This is expected, however, as the ion beam can suppress the $T_c$ slightly also in the proximity of the irradiated line due to the Gaussian beam shape. The Dynes broadening parameters for junctions JJ7 and JJ8 were $0.18 \Delta$ and $0.25 \Delta$. 

\begin{figure*}[t!]

\includegraphics[width=0.7\linewidth]{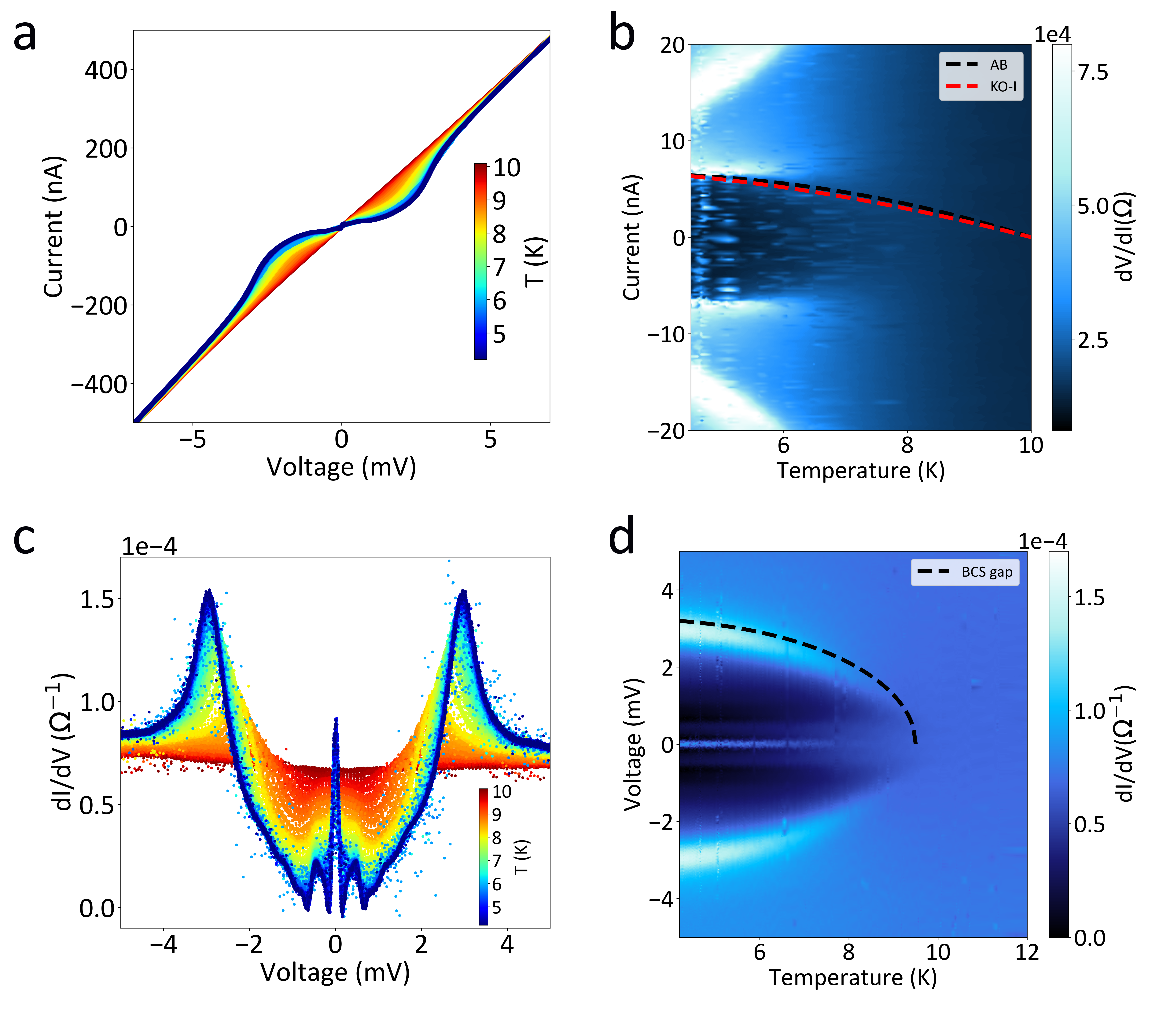}

\caption{DC transport measurements of Josephson junction JJ7 fabricated with high enough fluence to push the barrier to the insulating phase. The junction is clearly SIS-type, as can be seen from IVC (a) and dI/dV plots (c,d). Both Kulik and Omelyanchuk theory (KO-I) and Ambegaokar-Baratoff (AB) theory predictions for $I_c(T)$ are shown in the plot (b). Temperature dependence of the superconducting gap in (d) follows the BCS theory nicely with $\Delta \approx 1.5$\,meV. }

\label{fig:JJ7_ivc}
\end{figure*}

\subsubsection{Critical current as a function of temperature}

From the measured $dV/dI(T,I)$ maps (Figure \ref{fig_ivc}, bottom panels) we can extract the temperature dependent critical current $I_c(T)$ that can be used to determine the weak link properties. When a barrier is strong, $I_c$ has a weak temperature dependence at low temperatures, and this dependence gets weaker as the barrier gets stronger. This is expected, as $I_c(T)$ scales exponentially in the case of transparent diffusive metallic barriers, and the saturation at low temperatures is signature of either an insulating barrier, or a very strong diffusive barrier \cite{Likharev_weak_links}. We do not consider the clean limit here, as the barrier is highly disordered due to the irradiation. 

In order to get a more quantitative analysis of $I_c(T)$, we performed numerical modeling, assuming diffusive transport through the weak link. For $I_c(T)$ in junctions JJ1 and JJ2, a numerical simulation was performed using \textit{usadel1}-code \cite{Paulinkoodi} that solves the quasiclassical Usadel equations in the diffusive limit. Simulation fits using this code are shown as dashed lines in Figures \ref{fig_ivc} for junctions JJ1 and JJ2. As a fit parameter we used $\Delta/E_{th}$, where Thouless energy $E_{th}$ represents characteristic energy scale of the SNS junction, defined as $E_{th} = \hbar D/L^2$, where $D$ is diffusion constant and $L$ the junction length.

For junction JJ1, we found a good fit to $I_c(T)$ with $\Delta/E_{th} = 6$, so that we have $E_{th} \approx 240\,\mu$eV, from which we determine the diffusion constant to be $D \approx 0.7$\,cm$^2/$s, and thus from Eq. \ref{eq:ksii} the normal coherence length $\xi_N \sim 4.5$\,nm at 4.2\,K. 
This puts junction JJ1 well into the long junction limit $L > \xi_N$. However, an excess  current scaling factor $A = 1.8$, was needed to fit the data. 

The best fit to JJ2 $I_c(T)$ curve was achieved for $\Delta/E_{th} = 5$, so that $E_{th} \approx 250\,\mu$eV and $D \approx 3.4$\,cm$^2/$s. The Thouless energy $E_{th}$ is similar compared to JJ1, but as junction JJ2 is longer, the diffusion constant is larger. Also the fit was not as good as with JJ1: for JJ2 $I_c(T)$ it was possible to reproduce either the behavior at low temperatures or at high temperatures separately, but for the overall best fit, a compromise was needed, as shown in Fig. \ref{fig_ivc}. This is probably because by using Eq. \ref{eq:ksii} and $D$ determined from the fit, we get $\xi_N(4.2\,\text{K}) \approx $10\,nm and $\xi_N(1.5\,\text{K}) \approx $17\,nm , showing that the coherence length begins to approach the junction length (30 nm) at the low temperature end, causing most likely a crossover between the short junction limit to the long junction limit as a function of $T$. The scaling factor for $I_c$ in the fit was $A = 0.21$, so the maximum characteristic voltage would be five times larger. Interestingly, also for all other junctions fabricated with $W_{irr}\approx30\,$nm and $I_c\sim10-50 \mu$A, we always have a similar $I_c(T)$ behavior to JJ2, and we conclude that the $I_c(T)$ properties for JJ2 are common for these type of junctions, although not perfectly described with the theory presented here.

We used the \textit{usadel1}-code also for junctions JJ3-JJ8, but we found that Kulik and Omelyanchuk theory (KO-I)\cite{Kulik1975} for short weak links is applicable for these junctions, and the Usadel theory is not needed. KO-I theory was originally derived for diffusive one-dimensional wire, which applies when $\ell \ll L_j \ll \xi_N$. Within these assumptions, the supercurrent is given by
\begin{equation}
I(T,\varphi) = \frac{2\pi k_BT}{eR_n}\sum_{\omega_n}\frac{2\Delta\cos(\varphi/2)}{\Omega_n}\arctan\frac{\Delta\sin(\varphi/2)}{\Omega_n},
\label{eq:KO-I}
\end{equation}

where $\Omega_n^2 = (\hbar\omega_n)^2 + |\Delta|^2  \cos^2(\varphi/2)$, $\varphi$ is the phase difference, and Matsubara frequencies $\omega_n$ are defined such that $\hbar\omega_n = \pi k_B T (2n+1)$. This same exact form can also be derived from generalization of point contacts, valid for arbitrary level of disorder, when a large number of channels and diffusive limit $\ell \ll \xi_N$ is assumed \cite{Golubov2004}. Even more importantly, KO-I equation turns out to be valid for tunnel junctions with uniform spatial distribution of localized states \cite{Naveh2000,Golubov2004}.
As previously mentioned, the insulating phase induced by disorder has localized states, and it is possible that $I_c(T)$ follows KO-I theory even in the insulating side of the transition.

To determine $I_c(T)$, we solved the maximum supercurrent from Eq. \ref{eq:KO-I} numerically by summing the Matsubara frequencies up to $n = 10^5$. The superconducting gap $\Delta$ was calculated with the BCS temperature dependence \cite{tinkham_book} and using $\Delta(T=0) = 2k_BT_c$ that has been shown to be more suitable for nitrides. 
Results of these calculations are shown in Figure \ref{fig_ivc} (bottom panel) as dashed lines with the label KO-I.

In the case of junctions JJ1 and JJ2, it is clear that the approximation for a short junction is not fulfilled and KO-I theory can't be used. However when $J_c$ of a junction is decreased, we  have increasingly better fits to KO-I theory.
For JJ3, we can get an approximate fit to KO-I theory with a prefactor $A=0.45$, but there is some deviation between $I_c(T)$ and KO-I theory (not shown). When \textit{usadel1}-code is used, we see that the KO-I limit of the code $\Delta/E_{th} \ll 1$ fits $I_c(T)$ best, indicating that the full Usadel theory does not explain this discrepancy any better that KO-I theory.

For junctions JJ4-JJ6, the fits to KO-I theory are very good, suggesting that these junctions are in limits where KO-I theory is applicable. For JJ4, best fit is achieved with a prefactor $A = 0.35$ (Fig. \ref{fig_ivc}) and a 12\,nm junction length, showing  
only a small deviation between the KO-I theory and the measured $I_c(T)$. For JJ5 we get a good fit with a prefactor $A=0.7$, and the deviation between measured the $I_c(T)$ and KO-I theory begins to be negligible.

For JJ6, we also compare the data to the Ambegaokar-Baratoff (AB) theory\cite{Ambegaokar1963}
\begin{equation}
I_c(T) = \frac{\pi \Delta (T)}{2eR_n}\tanh(\frac{\Delta(T)}{2k_BT}),
\label{eq:AB}
\end{equation}
valid for SIS tunnel junctions without localized states in the barrier. We observe that the AB fit for JJ6 (black dashed line in Fig.\ref{fig_ivc}) is not as good as with the KO-I theory. As the KO-I theory fits almost perfectly to the measured $I_c(T)$s of junctions JJ5 and JJ6, 
these junctions must have either a diffusive normal metal barrier or an insulating barrier with localized states. For JJ6, the purely theoretical $I_cR_n$ value was significantly higher than the measured one, where in fits we used the prefactor $A = 0.14$ for the KO-I theory and $A = 0.18$ for the AB theory to get good fits. 

For junctions JJ7 and JJ8 it is difficult to fit theoretical predictions to the measured $I_c(T)$as $I_c$ is suppressed and thermal noise affects measurements. However, for junction JJ7 we have nevertheless fitted both AB and KO-I predictions to the measured $I_c(T)$, and both seem to fit data rather well (Figure \ref{fig:JJ7_ivc} (b)). 

\input{Table1_v8_vertical}

\subsubsection{Characteristic voltages}

The characteristic voltage $I_cR_n$ values were also determined for each junction. All SNS-junctions feature relatively high $I_cR_n$ values, ranging between $0.2-1.7$\,mV. From the RCSJ fits it can be seen that an excess current contribution in these junctions is mostly small,
 and for example for junction JJ2, the excess current contribution is essentially zero. This can been seen both from the RCSJ fits and the magnetotransport measurements (Figure \ref{fig_icb_map}), where the critical current is completely suppressed at the nodes of the Fraunhofer-pattern. Junctions with the highest $I_cR_n$ are JJ3-JJ5, which have a characteristic voltages of $\sim$\,1.5\,mV. One explanation for the high values is that the sheet resistance of these junctions is quite close to $R_q$, and there is evidence that $I_cR_n$ values for weak links closer to the phase transition are higher\cite{Yu_2006}. Because of some nonlinearity in the IVC, it is difficult to estimate the excess current precisely, but based on the RCSJ fits, its contribution is below 10\% for JJ4. Hence the high $I_cR_n$ is not caused by an excess current contribution. For the case of JJ3 and JJ5, there are strong signatures of multiple Andreev reflections (MAR) 
 in the IV-characteristics, and thus there is some excess current carried by MAR. For JJ5 we estimate the excess current to be $\sim 0.6$\,$\mu$A, yielding corrected $I_cR_n \approx 1.4$\,mV. For junction JJ3 the contribution is smaller,
 below 10\%. The achieved $I_cR_n$ values are already suitable for example for RSFQ applications, and with some tuning it should be possible to push these characteristic voltage values even higher.


\begin{figure*}
\includegraphics[width=\linewidth]{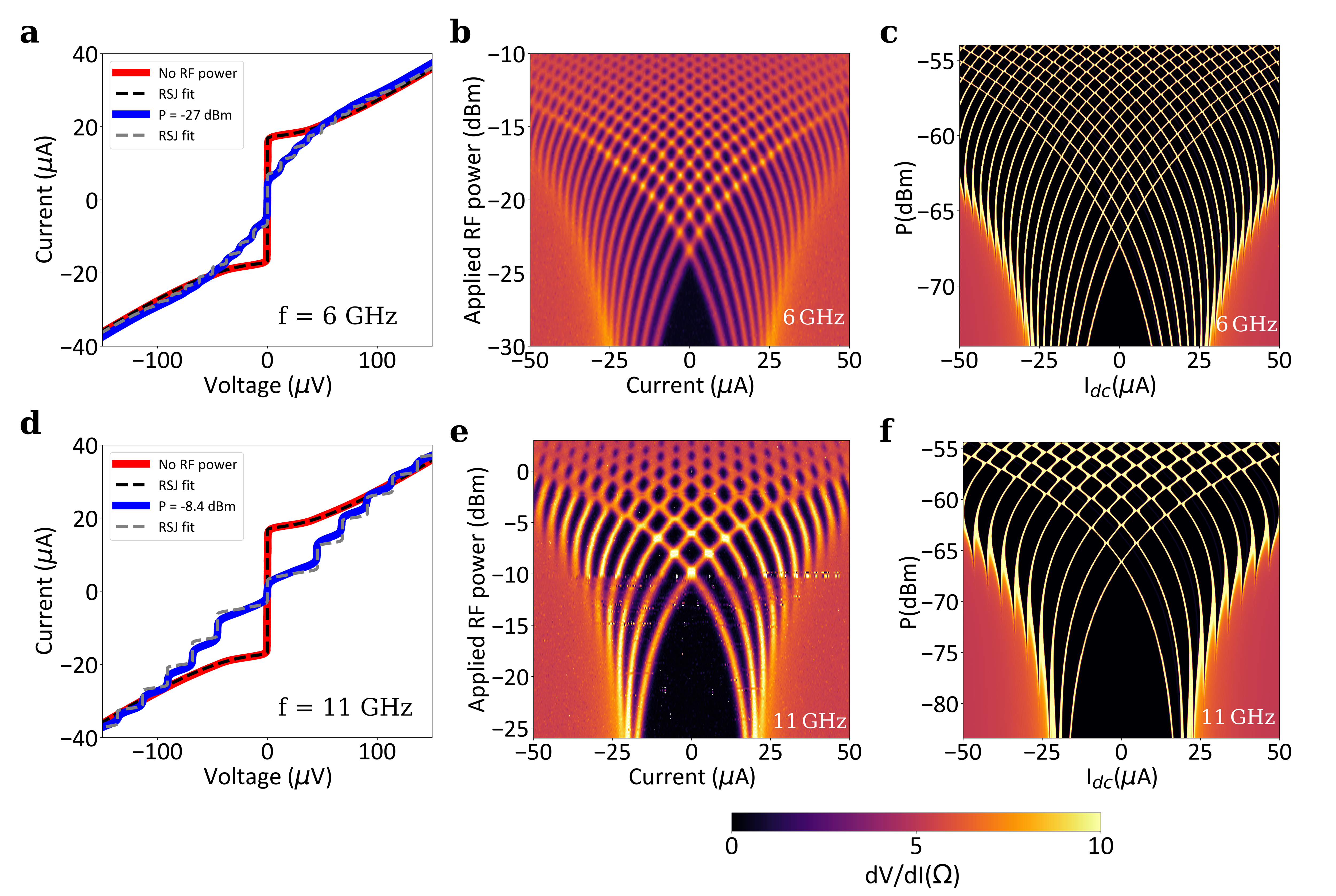}

\caption{RF characterization of junction JJ2, showing both 6 GHz (a,b) and 11 GHz (d,e) measurements and comparison to simulations (c,f). Panels (a) and (d) show selected IV curves with and without RF power, and theoretical fits using the RSJ model. Panels (b) and (e) show the so called Shapiro map, i.e. the measured $dI/dV$ as a function of both RF power and DC bias current. As a comparison, full Shapiro map simulations are shown in (c) and (f). The best fit parameters were found to be $R_N = 4.8\,\Omega$, $Ic = 17.5\,\mu$A.}
\label{fig_RF}
\end{figure*}

\begin{figure*}
		\includegraphics[width=0.9\linewidth]{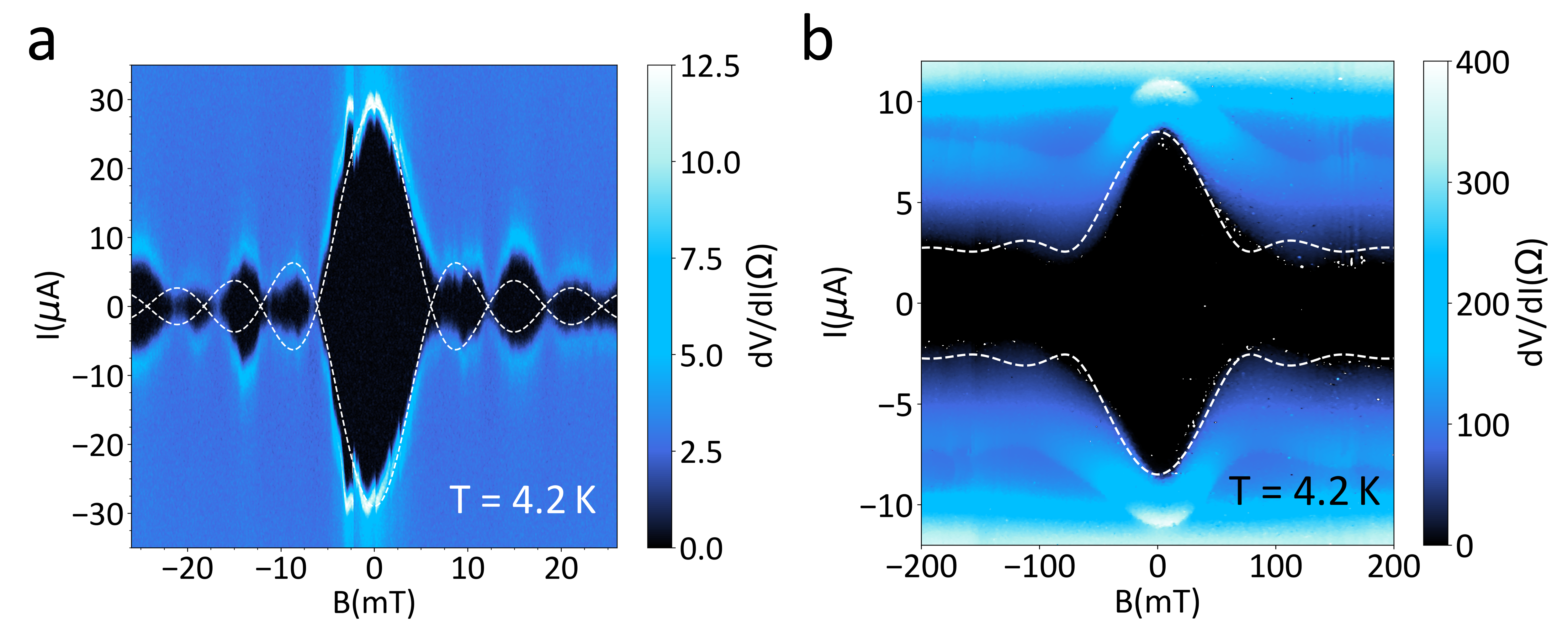}
		\caption{Panel (a) shows the differential conductance of junction JJ2, measured as a function of magnetic field at 4.2\,K with a fit to Eq. \ref{eq:fraunhofer}(white dotted line). There is clearly a periodic modulation of the critical current as a function of the magnetic field, and at the nodes the critical current vanishes completely. Panel (b) shows the same for junction JJ4, and a fit to Eq.\ref{eq:yanson}. In this junction there is not as clear modulation as a function of magnetic field.} 
		\label{fig_icb_map}
\end{figure*}

\subsubsection{Microwave measurements}
In order to characterize the junctions further, we performed electrical measurements of JJ2 under microwave irradiation. We used 2.1-11 GHz RF radiation produced with an Anritsu 68367C signal generator. The minimum signal output used was -10 dBm, but to get a lower power output for the measurement we used a 20 dBm attenuator in the RF line. The RF signal was carried into the cryostat via a low loss coaxial cable, and the RF radiation was coupled to the sample with a simple loop antenna. The microwave measurements were carried out by scanning the RF output power with a fixed frequency, and measuring $dV/dI$ directly using a four wire lock-in technique at each power level.

For a theoretical comparison, we performed RCSJ simulations for the IV curves, and found that at 4.2\,K, the junction behavior follows reasonably well the simpler RSJ model \cite{tinkham_book}, which was used to simulate the RF modulated response of the junction vs. current and RF power (Shapiro map). 

Figs. \ref{fig_RF} (a) and (d) first shows examples of IVCs of junction JJ2 under 6 GHz and 11 GHz RF illumination, respectively, demonstrating clearly Shapiro steps and good agreement with the RSJ model simulations. The corresponding full Shapiro maps are shown in Figs. \ref{fig_RF} (b) and (e),  showing nearly ideal behavior. Comparison to the RSJ model simulated Shapiro maps (Fig. \ref{fig_RF} (c) and (f)) shows very little deviation between the experiment and the simulation, indicating a high quality junction. The Shapiro steps are visible up to high order of $> 20$ for the 6 GHz illumination.

\subsubsection{Magnetic field measurements}
Magnetotransport measurements were performed at 4.2\,K using a liquid helium dipstick and a 4\,T superconducting magnet, producing a magnetic field perpendicular to the chip. 
The measurements were performed by scanning the coil current and measuring $dV/dI$ directly using a four wire lock-in technique. For a rectangular junction geometry, the critical current is expected ideally to follow \cite{tinkham_book}
\begin{equation}
I_c(B_z) = I_{c0} \left| \frac{\sin(\pi B_z A/ \phi _0)}{\pi B_z A/\phi _0 } \right|,
\label{eq:fraunhofer}
\end{equation}

where the junction area perpendicular to the B-field is $A = Lw$ and the magnetic field period for which the phase changes by $2\pi $ is thus $\Delta B = \phi_0/A$. For the length of the junction, the Josephson penetration depth $\lambda_j$ has been taken into account, so that $L = L_j + 2\lambda_j$ \cite{tinkham_book}. In Figure \ref{fig_icb_map}, we present the results of these magnetotransport measurements for two junctions, JJ2 and JJ4. Junction JJ2, made to a \SI{2}{\micro\meter} wide superconducting wire has clearly a sinusoidal modulation with a vanishing critical current at the nodes. This proves that a Josephson current flows through the junction, although the pattern slightly differs from the expected $\sin(x)/x$ behaviour. Nodes are roughly equally spaced, but peculiarly, their height does not decrease monotonically as the magnetic field increases. It is known that while Equation \ref{eq:fraunhofer} holds well for uniform current distribution and for wide ($w \gg \lambda_j$) and thick films, in planar junctions made from very thin films ($\lambda_j \gg t$) this diffraction pattern is often altered because of flux focusing effects\cite{Rosenthal1991}. For very narrow junctions, the usual magnetic period is replaced by $\Delta B\approx 1.84 \Phi_0/w^2$, as was shown by Rosenthal et al. \cite{Rosenthal1991}. These flux focusing effects can cause also aperiodic node spacing, as well as nonmonotonically decreasing node height \cite{Suominen2017}, as observed for JJ2. For JJ2, the observed period is $\Delta B \approx 5.8$\,mT, and if we calculate the Josephson penetration depth using this fact and Eq. \ref{eq:fraunhofer}, we get 75$\pm$ 10\,nm. This differs considerably from the usual $\sim$200\,nm reported in literature for NbTiN \cite{Lei2005,Yamamori2010}. The BCS prediction for the London penetration depth is given by
\begin{equation}
\lambda \approx 105\,\text{nm} \sqrt{\frac{\rho(\mu\Omega\text{cm})}{T_c(K)}},
\label{eq:penetration_length}
\end{equation}
and using this we get $\lambda \approx 150$\,nm for our best films, still a factor of two larger. However, flux focusing effects \textit{increase} the effective length, and thus are not able to account for the discrepancy. Consequently, for junction JJ2 the Rosenthal correction for flux focusing gives a way too short node spacing ($\sim 1$\,mT).

The magnetotransport measurements of junction JJ4, made from a 300\,nm wide wire, did not show the expected Fraunhofer diffraction pattern, Eq. \ref{eq:fraunhofer}. The critical current as a function of the applied magnetic field was not suppressed completely at the nodes, and the shape of $I_c(B)$ curve shows additional constant residual $I_c$ on top of the diffraction pattern (Fig.\ref{eq:fraunhofer} (b)). We surmise that this persisting $I_c$ may be caused by random structural inhomogenities to which Josephson vortices are pinned \cite{Yanson1970,tinkham_book}. Yanson \cite{Yanson1970} showed that the shape of $I_c(B)$ is then altered to
\begin{equation}
I_c(B) = I_{c0}\sqrt{(1-\gamma^2)\left(\sin(X)/X\right)^2+\gamma^2},
\label{eq:yanson}
\end{equation}
where $X = \pi B/B_0$. Structural inhomogenities are introduced to this equation by a random additional current $I_1(x)$ that is due to inhomogenities in the barrier, and by a factor $N$ that represents the amount of structural inhomogeneity so that $\gamma^2 = (\overline{I_1^2}/I_0^2)\cdot(1/\pi N)$. When we fitted Eq. \ref{eq:yanson} to the measured $I_c(B)$ of JJ4, we got a reasonable fit with $\gamma = 0.3$. From Figure \ref{fig_icb_map} (b), we can determine the position of the first minimum, and from this we get $\Delta B \approx 90$\,mT. When we use Eq. \ref{eq:fraunhofer} to calculate the Josephson penetration depth for JJ4 we get 40$\pm$ 20\,nm. Hence, the determined $\lambda_j$ for both junctions are reasonably close to each other, but differ considerably from the literature values as well as the BCS prediction. For JJ4, the Rosenthal correction gives $\Delta B \approx 60$\,mT, which is not too far from the measured value and thus might explain the observed node spacing in this case.




\section{Conclusions}
In this study, we have shown for the first time a precise local control of superconductivity in NbTiN films using He ion irradiation, and shown that this can be utilized for the fabrication of highly tunable weak links for Josephson junctions. We also  characterized the fabricated Josephson junctions extensively using both dc, microwave and magnetotransport measurements, 
and show that the fabricated junctions are of high quality. Additionally, we showed that we are able to tune the barrier strength and fabricate weak links spanning from strongly metallic to insulating limit. 
The junctions with strongest barrier have excellent thermal stability (weak dependence on temperature) 
as high as at 1.5\,K, which is quite promising for example for higher temperature quantum information and superconducting electronics applications. %

 In addition to this, we have pushed the barrier to the insulating side of the SIT to create SIS junctions. Interestingly, these junctions have relatively 
low ($<100$\,meV) tunneling barriers. As this is parameter region that is not often reached, there might be several interesting applications for the directly written SIS junctions. As NbTiN is also excellent for microwave applications, we see that a promising avenue to explore is the 
 fabrication of complete Josephson junction quantum devices (qubits, for example) with the help of direct writing,
using only a single superconducting film. This allows resonators and junctions to be the same material, possibly eliminating some of the sources of noise and decoherence associated with extra interfaces and materials.

%


\section{Acknowledgements}

\begin{acknowledgments}
The authors wish to thank Geng Zhuoran and Pauli Virtanen for discussions and providing parts of the codes, as well as Antti Kanniainen for the help with the RF measurements. We also thank Tatu Korkiam\"aki for the help with some of the cryogenic measurements.

This research was supported by the Research Council of Finland projects number 341823 and 358877 (the Finnish Quantum Flagship), the Vilho, Yrj\"o and Kalle V\"ais\"al\"a Foundation of the Finnish Academy of Science and Letters and Jenny and Antti Wihuri Foundation.
\end{acknowledgments}

\bibliography{references.bib}
\end{document}


\thispagestyle{empty}
	\vspace{2cm}
	\begin{center} 
		{\LARGE Supplementary Information for Highly tunable NbTiN Josephson junctions fabricated with focused ion beam}\\
		\vspace{1cm}
		Aki Ruhtinas$^1$ and Ilari Maasilta$^1$ \\
		\vspace{0.2cm}
		$^1$Nanoscience Center, Department of Physics, University of Jyv{\"a}skyl{\"a}, FI-40014 Jyv{\"a}skyl{\"a}, Finland
	\end{center}
   \vspace{5cm}

\section{Fabrication details}
\subsection{Pulsed laser deposition}
For NbTiN thin film deposition we used reactive pulsed laser deposition setup \cite{Torgovkin_2018,Chaudhuri2013,Chaudhuri2011} equipped with a Q-switched Nd:YAG laser (Ekspla NL301 HT) operating at its fundamental 1064\,nm wavelength. The laser fluence at the target was $\sim$6\,J/cm$^2$ and the pulse repetition frequency 10\,Hz.  As a target, we used two 99.99\% purity NbTi alloy targets (Matsurf Technologies Inc.) with compositions Nb$_{0.39}$Ti$_{0.61}$ and Nb$_{0.5}$Ti$_{0.5}$. The target was rotated and laser pulses were sweeped over the target with optics in order to reduce splashing of the target. With this setup, we find that films are very smooth and free of droplets. The target was ablated in ultra-high purity (6N purity) nitrogen atmosphere using $\sim 50$\,mTorr pressure. As a substrate, we used cubic (100)-oriented MgO made by Crystec GmbH. The substrate temperature was kept at $700\,^{\circ}$C, and the substrates were glued to substrate heater using silver glue in order to maximize the thermal contact. The growth rate of the film was $\sim$1\,nm/min. 

\subsection{Electron beam lithography and plasma etching}
Bonding pads and superconducting wires with widths ranging between 0.2-2\SI{}{\micro\meter} were defined to the PLD deposited nitride film with electron beam lithography and reactive ion etching. Before the lithography, the chip was cleaned in isopropyl alcohol (IPA) bath using an  ultrasonic cleaner (Finnsonic) to get rid off the silver glue residues and other contaminants. Then,  a resist layer was spun on the samples at 4000\,rpm for 60\,s, with the resulting resist layer thickness approximately 400 nm. Here we used the negative e-beam resist \textit{AR-N 7520.17 new} because it has excellent plasma etch resistance and because a negative resist is more suitable for our device geometry.  After spinning, an e-beam lithography tool (Raith E-Line) was used to expose the resist to define the pattern, with a 110-\SI{150}{\micro\coulomb/cm^2} dose depending on the pattern size. After the exposure, the resist was developed in undiluted AZ 351B developer for 50s, and rinsed with IPA, and reactive ion etching (RIE) was performed (Oxford Plasmalab 80+ RIE) to remove the NbTiN film outside the wire and bonding pad regions, using 60\,W RF power, 100,sccm SF$_6$ flow, and 5 sccm O$_2$ flow at a pressure of 70\,mTorr. This results in etching speed of roughly 15\,nm/min. Finally, the remaining resist was removed using O$_2$ plasma cleaning in the same RIE tool.

\section{Helium ion irradiation and milling}
Josephson junctions were defined to the superconducting wire using a helium ion beam irradiation. We used $30\,$kV He$^+$ beam in a Zeiss Orion Nanofab helium ion microscope with varying beam parameters. The irradiation was performed with a 1\,nm step size and typically a \SI{1000}{\micro\second} dwell time for each step, with the number of repeats adjusted to get the desired fluence. Patterning was performed with built in pattern generator and ZEN software.
We used mostly \SI{10}{\micro\meter} aperture and spot size 4, resulting in a beam current of 0.5-1\,pA. In order to obtain smaller beam size, we used spot size of 5-6. With this, we reached $\sim$3\,nm linewidth in best cases. For larger linewidth irradiations we used spot sizes up to 0, resulting to beam current of $\sim$20\,pA. In some cases, beam was defocused (with 1-3\,V offset) to get wider linewidth. With HIM, increasing beam size often increases stability and thus more reproducible fabrication is possible. Even though the ion beam current of the instrument is not as stable as an electron beam in a SEM and fluctuates over time, when beam and trimer is correctly tuned, current is stable enough for reproducible junction fabrication.\cite{Barlow2016}

\subsection{Helium ion fluence calculation}
\begin{figure}[H]
	\begin{center}
		\includegraphics[width=0.48\linewidth]{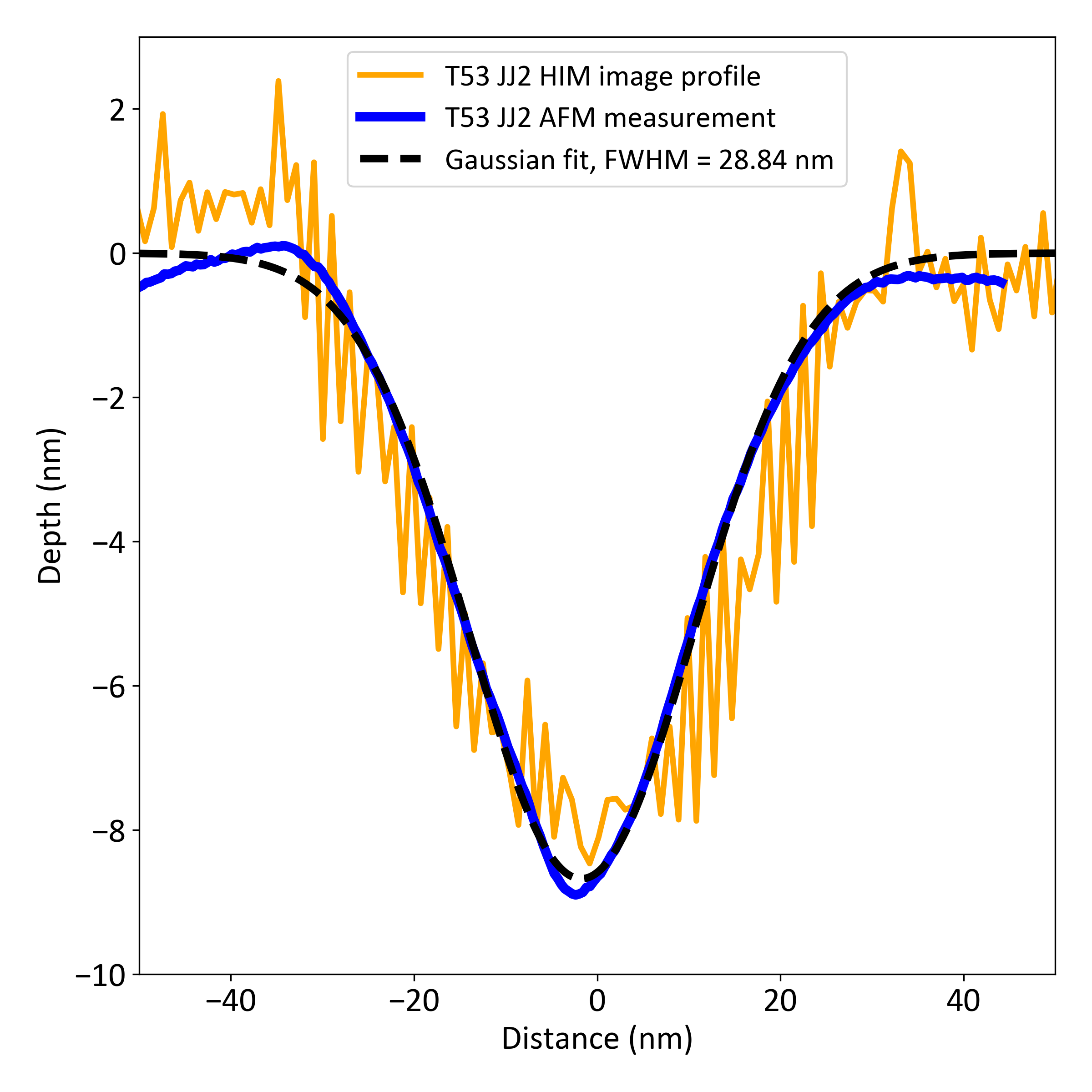}
		\includegraphics[width=0.48\linewidth]{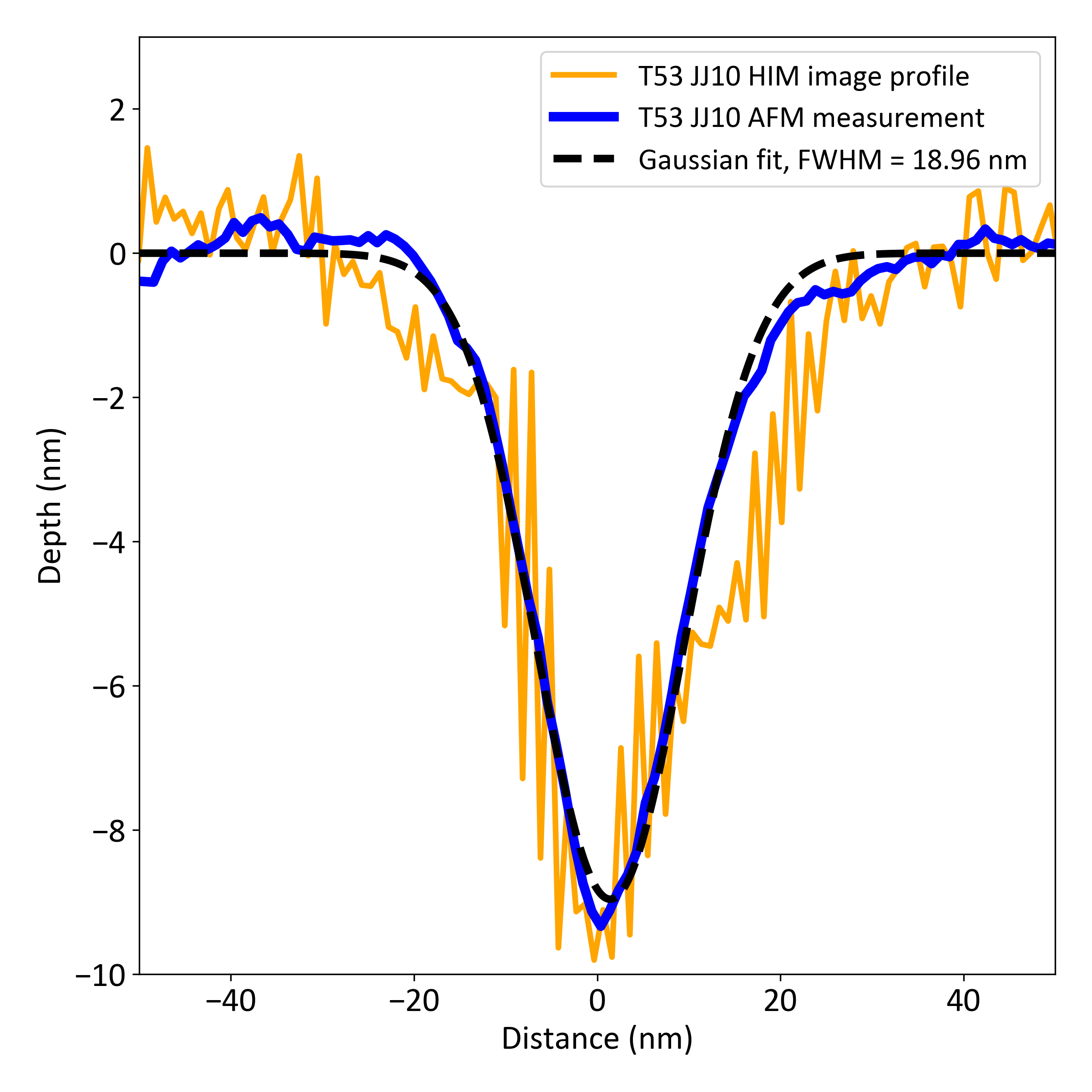}
		\caption{HIM image line profile plotted against measured AFM profile of the junction and gaussian fit to AFM data. Both methods agree well, especially when only darkest region of HIM image is fitted. HIM image line profile grey values are scaled to AFM measurement.}
		\label{fig:him_afm_compare}
	\end{center}
\end{figure}

While software used in HIM irradiation calculates fluence, this value can be almost order of magnitude wrong. Instead we calculated fluence based on averaged helium ion current $I_{HIM}$, patterning parameters and realized irradiation width. Irradiation width $w_{irr}$ was characterized as FWHM value of the image line profile. In all irradiations we used pixel size $\Delta x = $1\,nm, dwell time $t_d = $1\,ms with varying number of line repetitions $N_{repeat}$. Now fluence is given by 
\begin{equation}
\Phi = \frac{I_{HIM}\cdot t_d \cdot N_{repeats}}{e \cdot w_{irr} \cdot \Delta x\cdot10^{-4}},
\end{equation}
where $e$ is elementary charge and all input variables are in SI units and thus $\Phi$ is in units of ion/cm$^2$. While standard deviation of $I_{HIM}$ is order of 1-2\%, more substantial uncertainty comes from determination of $w_{irr}$. In most cases, this was estimated from HIM images of the irradiated line, and especially  narrow junctions had large width uncertainties using this method. However we have made comparisons of this method against AFM measurements of irradiated lines (Figure \ref{fig:him_afm_compare}), and in most cases these two methods agree within 1-2 nanometers. Often only quick HIM imaging of the junctions is possible in order to not irradiate junction further, and thus images tend to be sligtly blurry resulting in larger determined irradiation widths compared to AFM. In addition to this, in HIM images gray value does not necessarily correspond to height, and thus this method needs to be used carefully. However, generally irradiation width determined from HIM image is a very good approximation of true width.

\subsection{SRIM simulations and helium ion milling rate estimate}
\begin{figure}[H]
	\begin{center}
		\includegraphics[width=0.6\linewidth]{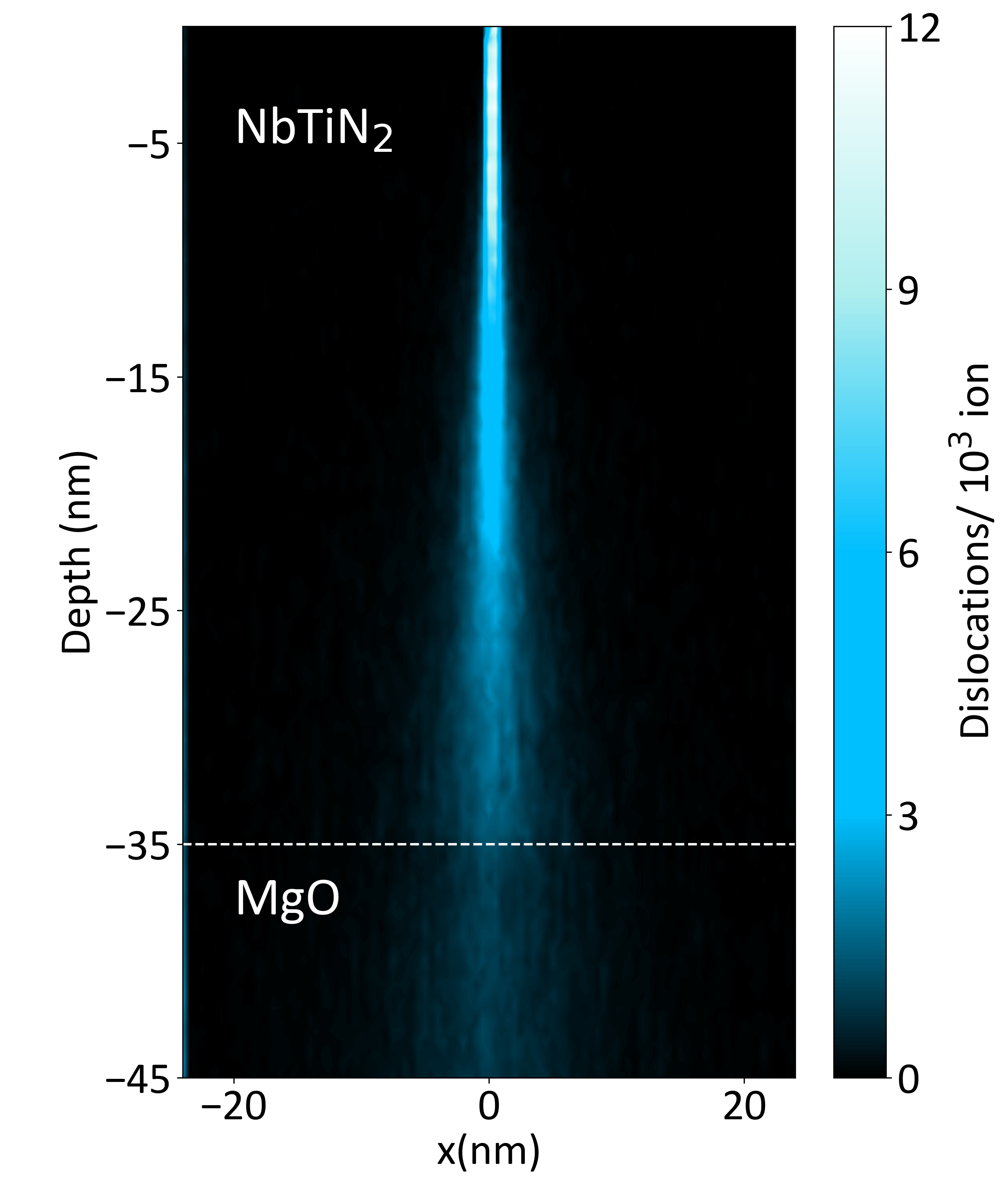}
		\caption{SRIM simulation of $30\,$kV He$^+$ ions impinging to 35\,nm NbTiN film on top of the MgO substrate.}
		\label{fig:srim}
	\end{center}
\end{figure}

In order to simulate He ions impinging to NbTiN film we have used SRIM software \cite{ZIEGLER2010}. For this, we set up 35\,nm NbTiN film on top of 500\,nm MgO substrate, and simulate $30\,$kV He$^+$ ions impinging to these layers. For simulations, we use 7.6\,g/cm$^3$ density for NbTiN \cite{Marek2013} and 3.58\,g/cm$^3$ for MgO \cite[p.108]{Ropp2013book} and assume chemical composition Nb$_1$Ti$_1$N$_2$. Usual starting value for surface binding energy (needed for accurate surface sputtering rate) is heat of sublimation, but for NbTiN we did not find literature value for this. Thus we approximated NbTiN surface binding energy with heat of sublimation of TiN, which is 818\,kJ/mole ($\sim$8.5 eV/atom)\cite{NASATND5027}. We simulated $\sim$ 25000 ions with these parameters, and found that ion range is roughly 144\,nm and sputtering yield $\sim$0.010\,atoms/ion for both Ti and Nb. For N sputtering yield was $\sim$0.016 atoms/ion, meaning that milled surfaces can be slightly nitrogen deficient. However, this should be only surface effect and thus shouldn't change chemical composition inside the film. Results of the simulation is shown in Figure \ref{fig:srim}. True value of the disorder is probably lower than simulated because at room temperature, substantial amount of target damage is usually efficiently repaired by recrystallization. This effect has been demonstrated with both experiments \cite{Pelaz2004} and simulations \cite{Nord2002}, and it is most prominent in metals, so a large deviation from SRIM simulations is expected. In addition, SRIM assumes a perfect crystal for each impinging ion, and does not take into account milling and the accumulation of dislocations. Nevertheless, dimensions of the irradiated region should be accurate. 

\begin{figure}[H]
	\begin{center}
		\includegraphics[width=0.6\linewidth]{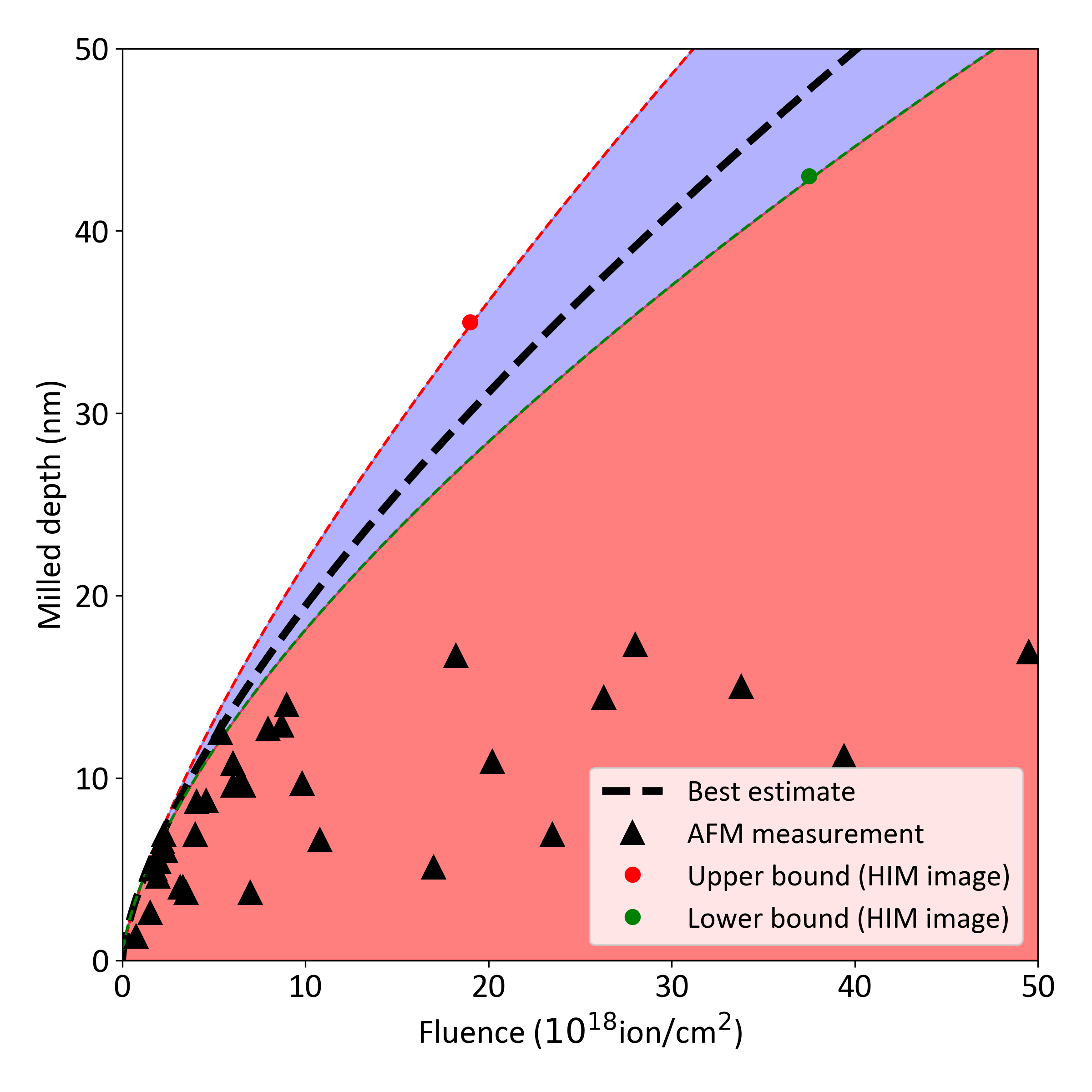}
		\caption{Experimentally determined milling depth as a function of helium ion fluence}
		\label{fig:milling_plot}
	\end{center}
\end{figure}

For the milling calculations, NbTiN we used molar mass of $168.787$\,g/mol and literature value 7.6\,g/cm$^3$ \cite{Marek2013} for the density, we get NbTiN$_2$ number density $ n_{\text{NbTiN}} = 2.7\cdot10^{22}$ atom/cm$^3$. If we ignore differences between elements and assume some effective sputtering yield $Y$, we get equation for milling as a function of fluence $\Phi$:
\begin{equation}
d_{mill}[\text{cm}] = \frac{Y [\text{atom/ion}] \cdot \Phi[\text{ion/cm}^2]}{n_{\text{NbTiN}}[\text{atom/cm}^3]}.
\label{eq:simple_mill}
\end{equation}
With simulated sputtering rates, we expect that 35\,nm film should be milled down near fluence of $\sim 1\cdot10^{19}$ He$^+$/cm$^2$ which is less than half of the experimentally determined value (>$2\cdot10^{19}$ He$^+$/cm$^2$). However, it is known that sputtering yield decreases as trench depth increases and that sputtering yield is nonlinear function of fluence \cite{Fowley_2013}. There are many processes affecting this, but in high fluences redeposition begins to be dominating. Redeposition is stronger when milled trench is deeper and narrower, and thus there is upper limit in trench depth. In order to quantify this, we performed milling experiments with HIM. In order to estimate this, we have made AFM measurements of trench depth (Figure \ref{fig:milling_plot}). However, from this one can see that AFM measurements underestimate true trench depth because of finite tip dimensions. In order to estimate high aspect ratio milled trench depth, we have used HIM imaging with tilted stage. There are large uncertainties in this method, but we have been able to get some lower bound (green point). In addition to this, we performed milling tests and observed changes in charging effects of milled lines. As substrate is insulating, completely milling through sample  Because junction JJ0 made with highest fluence was still conducting, we conclude that there is still some amount of NbTiN film left. This is also backed by HIM images, from which NbTiN seems to be completely milled through when fluence is between fluences 20-30 $\cdot 10^{18}$ ion/cm$^2$. Similarly as in publication \cite{Fowley_2013}, we have used fitting function $Y = A\Phi^{B}$ in combination with Eq.\eqref{eq:simple_mill} to fit the experimental data. This fit was adjusted so that estimation when film is completely milled trough (based on HIM image charging as substrate is insulating) coincides with AFM measured data. Our best sputtering yield estimate is 
\begin{equation}
Y[\text{atom/ion}] = 0.011\cdot\left(\Phi[10^{18}\text{ion/cm}^2]\right)^{-0.32},
\end{equation}
so that when combined to Eq. \eqref{eq:simple_mill} our best milling estimate is
\begin{equation}
d_{mill}[\text{nm}] = 4.06\cdot\left(\Phi[10^{18}\text{ion/cm}^2]\right)^{0.68}.
\label{eq:mill}
\end{equation}
Using this, we get that 35\,nm film is completely milled through at fluence of $\sim 24\cdot 10^{18}$\,ion/cm$^2$ which seems to agree well with milling experiments. However, we should note that every data point has substantial uncertainty, and this milling function is only our best estimate as more accurate data is difficult to obtain.

\section{RCSJ simulations}
For Josephson junction DC and AC simulation we used RCSJ model, where current continuity equation is given by
\begin{equation}
\frac{\partial\phi(t)}{\partial t} = -\frac{1}{Q}\frac{\partial\phi(t)}{\partial t}-\sin(\phi(t))+i_{dc}+i_{ac}\cos(\Omega\tau),
\label{eq:RCSJ}
\end{equation}
where quality factor $Q$ is given by
\begin{equation}
Q = \omega_p R_n C,
\end{equation}
where $R_n$ is junction resistance, $C$ junction capacitance and plasma frequency $\omega_p$ defined as
\begin{equation}
\omega_p = \sqrt{\frac{2eI_c}{\hbar C}}.
\end{equation}
Here we have used reduced units, so that reduced frequency $\Omega$ is $2\pi f/\omega_p$ and reduced currents are $i_{dc} = I_{dc}/I_c $ and $i_{ac} = I_{ac}/I_c $. Equation \eqref{eq:RCSJ} was solved using in python numerical library \verb|scipy.integrate.odeint| and then $V_{dc}$ was calculated using average of $\frac{\partial\phi(t)}{\partial t}$ and Josephson relation. For simulations, we used  $5\cdot 10^{6}$ reduced time points with step size adjusted so that corresponding real time step would be $\sim 1\cdot 10^{-12}$\,s. When DC measurements were simulated, $I_{ac}$ was set to zero and simulation was performed with $\sim$ 100 $I_{dc}$ points. Simulation was then fitted against measured IVC to extract $R_n$ and $C$ of the junctions.

For Shapiro map simulations, same code was used but now 2000 $I_{dc}$ and 600 $I_{ac}$ points (in total $1.2\cdot 10^{6}$ points) were used to simulate Shapiro map accurately. $I_{ac}$ was then converted to power using equation
\begin{equation}
P_{dBm} = 10\log_{10}(1000\cdot(I_{ac}R_n/\sqrt{2})^2/R),
\end{equation}
where $R = 50\,\Omega$. As there is a lot of attenuation in cryostat lines and coupling to sample is not known, this power does not correspond to applied RF power


\section{Insulating state and variable range hopping theory fit}
\begin{figure}[H]
	\begin{center}
		\includegraphics[width=0.8\linewidth]{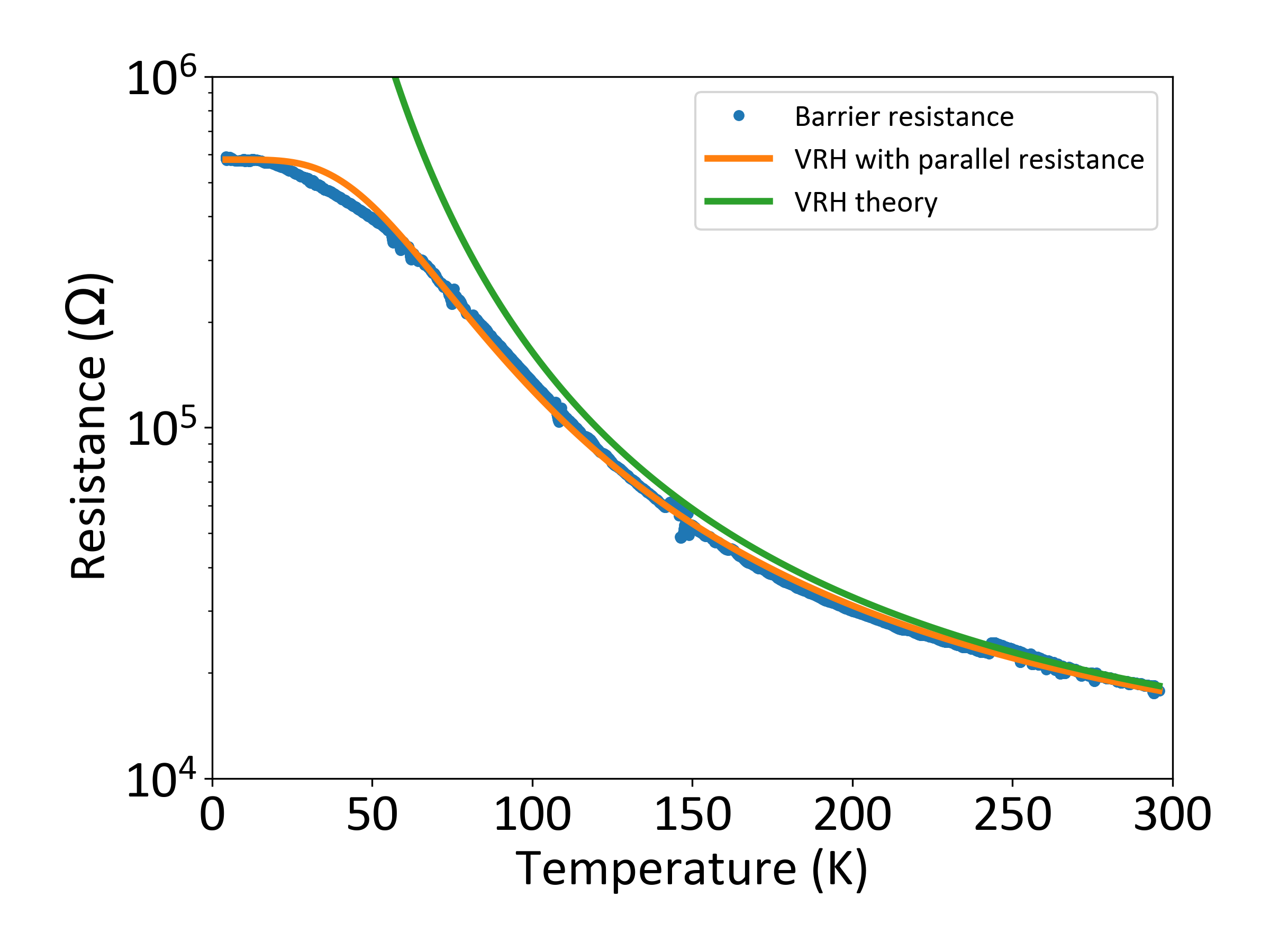}
		\caption{Variable range hopping theory fit to resistance as a function of temperature for sample J32LD5. Fit parameters are $\alpha \approx 0.56$ and $R_{parallel} = 580$\,k$\Omega$.}
		\label{fig:VRHfit}
	\end{center}
\end{figure}

With high enough helium ion fluence NbTiN can be pushed to insulating state. In insulating side of the transition, resistivity of NbTiN exhibits strong temperature dependence that is most commonly exponential. This exponential behaviour can be explained by variable range hopping (VRH) theories that assume localized electrons. When electrons are localized, conduction is due to states in narrow band near the Fermi level. Current carried by hopping process between localized states depends strongly on temperature, and Mott\cite{Mott1968} has showed that resistivity is of the form
\begin{equation}
\rho (T) = \rho_0 \exp(\frac{T_0}{T})^{\alpha},
\label{eq:MottVRH}
\end{equation}
where $\alpha = 1/4$ and coefficient $T_0$ is independent of temperature and can be used as a fit parameter. Mott's derivation assumes that density of states near the Fermi level is constant, and Efros and Shklovskii have worked out correction to this equation when electron-electron Coulomb interactions are taken into account \cite{Efros1975}. This correction changes only $\alpha$ to $1/2$, so that \eqref{eq:MottVRH} can be used for fitting when $\rho_0,T_0$ and $\alpha$ are left to fitting parameters. In addition to this exact R(T) behaviour is dictated by not only variable range hopping, but also possible tunneling currents as irradiated regions are quite narrow. In order to compensate this, we have added parallel resistance to VRH model. In Figure \ref{fig:VRHfit} is shown VRH theory fit to experimental data where $\alpha \approx 0.56$ and parallel resistance is $\sim$\,580\,k$\Omega$. VRH theory fits nicely experimental data, and $\alpha$ corresponds well with Efros-Skhlovskii theory, suggesting that electrons in NbTiN film are localized in the insulating side of the SIT.

\begin{figure}[H]
	\begin{center}
		\includegraphics[width=0.48\linewidth]{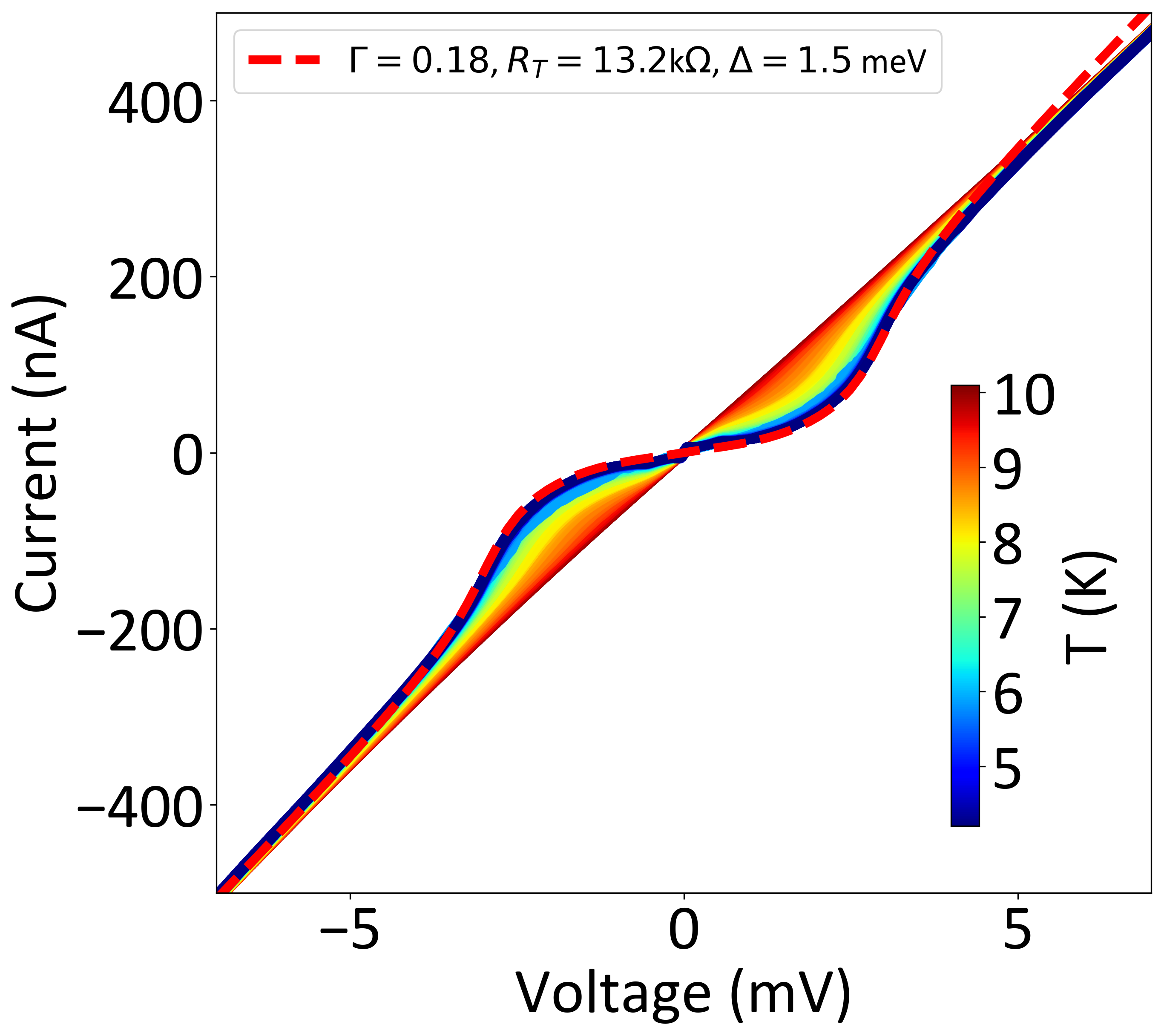}
		\includegraphics[width=0.48\linewidth]{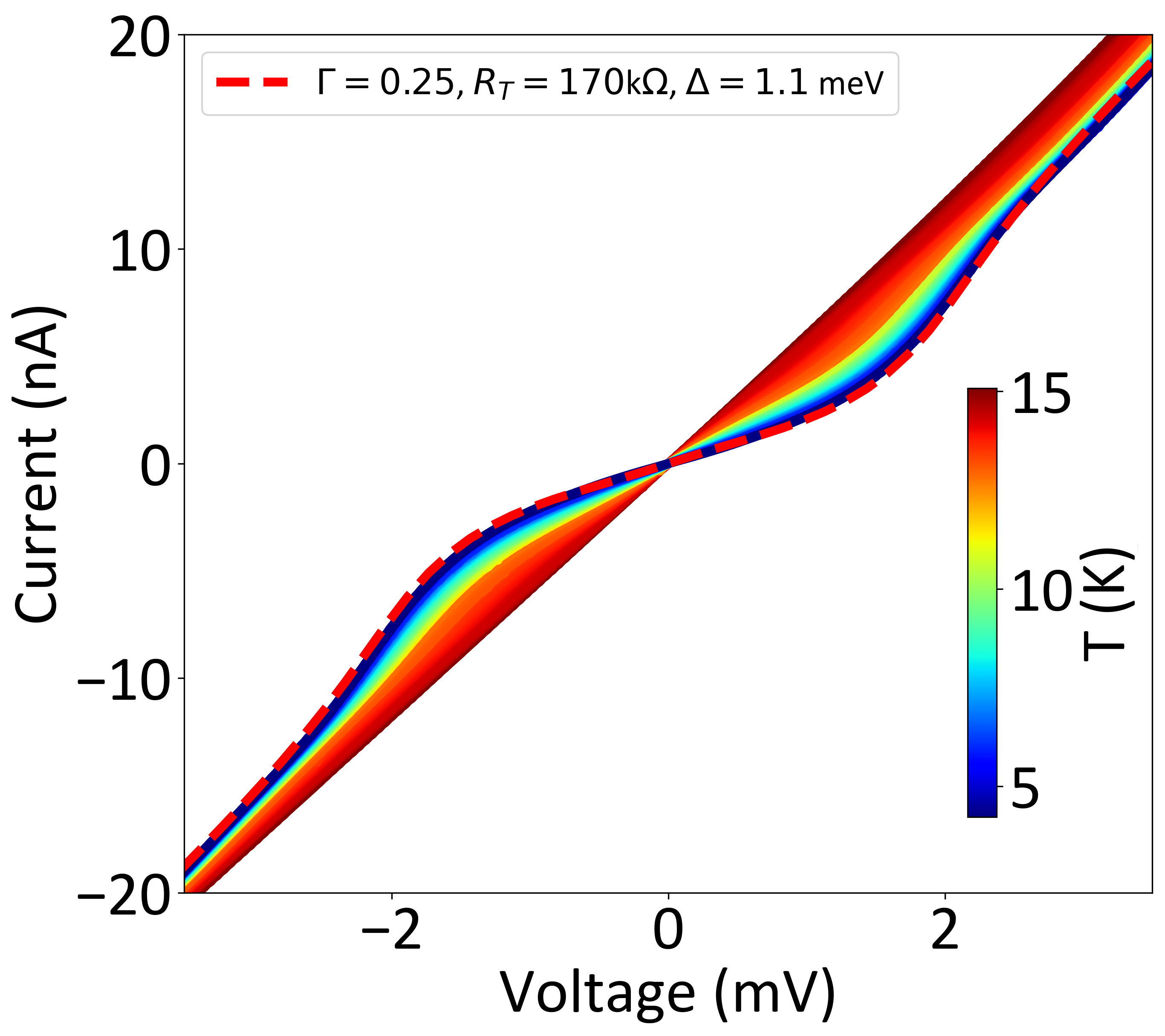}
		\caption{Quasiparticle tunnelnig model fits to junction JJ7 and JJ8 current-voltage characteristics.}
		\label{fig:sis_ivc_fit}
	\end{center}
\end{figure}

For SIS junctions JJ7 and JJ8 we also determined $R_n$, superconducting gap $\Delta$ and Dynes broadening parameter $\Gamma$ from fits to quasiparticle tunneling model \cite{tinkham_book}. These fits are shown in Figure \ref{fig:sis_ivc_fit}.

\bibliography{supplement_references.bib}

%% file: Table1_v8_vertical.tex
\newcommand{\notea}{\textsuperscript{a}}
\newcommand{\noteab}{\textsuperscript{a,b}}
\bgroup
\def\arraystretch{1.5}%
 \begin{table*}
  \caption{Junction parameters at 1.5\,K}
 \begin{tabular}{c c c c c c c c c}
\hline
   Name                                            &JJ1\notea       &JJ2            &JJ3                  &JJ4                    &JJ5\notea           &JJ6                &JJ7\noteab               &JJ8\noteab                                         \\
\hline                                            
\hline                                                                                                                                                                                                                           
   Lead $T_c$(K)                                   &12.84           &13.90          &14.2                 &15.70                  &14.2                &13.00              &14.2                     &13.80                             \\
   Film thickness (nm)	                           &35	            &48	            &35	                  &100	                  &35	               &35	               &35	                     &35	                                   \\
   Milled thickness (nm)	                       &26	            &27	            &16	                  &77	                  &13	               &16	               &7.7	                     &5.4	                           \\
   Junction width (nm)	                           &1500            &2030           &620                  &300                    &720                 &860                &660                      &1290                              \\
   Junction area $A_j$($\mu$m$^2$)                 &0.040           &0.055          &0.010                &0.023                  &0.009               &0.014              &0.0050                   &0.0069                            \\
   \hline                                                                                                                                                                                                            
   Fluence $\Phi$($10^{18}$He$^+$/cm$^2$)          &2.9             &11             &9.4                  &13                     &$\sim$12            &9.8                &17                       &19                                    \\
   Junction length $L_j$(nm)                       &14	            &30	            &7	                  &12	                  &3	               &3	               &3	                     &4                                     \\
   \hline                                                                                                                                                                                                            
   $I_c$($\mu$A)                                   &390             &44             &9.2                  &9.0                    &3.6                 &2.2                &0.15	                 &9$\cdot10^{-3}$                       \\
   $R_n$ ($\Omega$)                                &2.5	            &4.8	        &180	              &160                    &470                 &173	               &13.2$\cdot10^{3}$        &1.7$\cdot10^{5}$	                   \\
   $I_cR_n$(mV)    	                               &0.98	        &0.21	        &1.6	              &1.4	                  &1.7	               &0.38	           &2.0	                     &1.5	                       \\
   $J_c$(kA/cm$^2$)                                &978             &79.6           &90.6                 &39.1                   &38.5                &16.1               &2.96                     &0.13                              \\
   R$_{\square}$(k$\Omega$)	                       &0.27            &0.32           &15.5                 &4.00                   &110                 &49.6               &2900                     &5.5$\cdot10^{4}$                  \\
   $\rho$($\mu\Omega$cm)	                       &710             &880            &2.6$\cdot10^{4}$     &3.1$\cdot10^{4}$       &1.5$\cdot10^{5}$    &7.9$\cdot10^{4}$   &2.2$\cdot10^{6}$         &3.0$\cdot10^{8}$                \\
   \hline                                          
\multicolumn{9}{c}{\begin{footnotesize}\notea Parameters measured at 4.2\,K \textsuperscript{b} $I_c$ estimated from quasiparticle current inflection point\end{footnotesize}}
 \end{tabular}
 \end{table*}
\egroup